\documentclass[aps,showpacs,letterpaper,twocolumn,12point]{revtex4-1}
\usepackage{graphicx} 
\usepackage{amsmath}
\usepackage{color}      

\usepackage{amstext}
\usepackage{graphics}
\usepackage{amssymb}

\newcommand{\ket}[1]{|#1\rangle}

\begin{document}

\title{Analysis of a controlled phase gate using circular Rydberg states}

\author{T. Xia, X. L. Zhang, and M. Saffman}

\affiliation{Department of Physics, University of Wisconsin, 1150 University Avenue,
Madison, WI 53706}

\begin{abstract}

We propose and analyze the implementation of a two qubit quantum gate  using circular Rydberg states with maximum orbital angular momentum. The intrinsic quantum gate error  is limited by the finite Rydberg lifetime and finite Rydberg blockade shift. Circular states have  much longer radiative 
lifetimes than low orbital angular momentum states and are therefore candidates for high fidelity gate operations.  We analyze the dipole-dipole interaction of two circular state Rydberg atoms and present numerical simulations of quantum process tomography to find the intrinsic fidelity of a Rydberg blockade controlled phase gate.  Our analysis shows that the intrinsic gate error
can be less than  $9 \times10^{-
6}$ for circular Cs atoms in a cryogenic  environment.

\end{abstract}


\maketitle

\section{Introduction}

Highly excited Rydberg atoms are  promising candidates for quantum computing experiments, due to their long lifetime and strong interactions\cite{Jaksch2000,Saffman2010}. This strong, long-range and controllable interaction leads to the so-called Rydberg blockade effect in which  only one atom in an ensemble can be excited into a Rydberg state if the ensemble size is smaller than  the Rydberg blockade radius.
Using the Rydberg blockade effects, various schemes were proposed for fast quantum gates \cite{Jaksch2000,Lukin2001,Saffman2005b,Isenhower2011,Wu2010},
entangled state preparation \cite{Moller2008,*Muller2009,*Saffman2009b}, quantum algorithms \cite{AChen2011,Molmer2011}, 
quantum simulators \cite{Weimer2010,*Weimer2011}, and efficient quantum repeaters \cite{Han2010, *Zhao2010}.
Rydberg blockade, the central ingredient of the above schemes, has been demonstrated between two individual neutral atoms held in optical traps \cite{Urban2009,Gaetan2009},  and was used to demonstrate a two-qubit controlled NOT gate 
and entangled Bell states with fidelity of about $0.58-0.75$ after atom loss correction \cite{Isenhower2010,Wilk2010,Gaetan2010, Zhang2010}.

It is possible to estimate the fidelity error of a Rydberg blockade entangling gate from the atomic physics of the states used 
for Rydberg blockade\cite{Jaksch2000,Saffman2005a,Saffman2010}. 
The essential intrinsic errors are the finite lifetime of Rydberg states and the finite strength of the Rydberg-Rydberg blockade interaction.
 A rigorous fidelity measure for the gate operation can be found from   numerical integration of the master equation describing the gate evolution using real atomic parameters. The master equation solutions are then used to simulate quantum process tomography from which the gate process fidelity can be extracted. Using this approach we have shown that with low angular momentum $ns, np,$ or $nd$ states 
it is in principle possible to reach quantum process errors of $2\times 10^{-3}$ for both Rb and Cs atoms \cite{XZhang2012}. An error slightly less than $1\times 10^{-3}$ is projected for cryogenic operation at 4K, due to the increase in Rydberg lifetime.  
While these results are promising it is desirable for scalable implementation of fault-tolerant quantum computing architectures to reach gate errors
that are as small as possible. As the requirement for fault tolerance is strongly architecture dependent\cite{Aliferis2006,Aliferis2009,Fowler2009}
there is no precise requirement for the gate error. Nevertheless in order to avoid a blow up in the number of qubits needed for implementation the gate error should be well below the theoretical threshold and gate errors in the range of  $10^{-4}$ may be necessary for realization of concatenated code based error correction.

In this paper we propose implementing the two qubit Rydberg blockade  using high angular momentum circular Rydberg states 
$|m|=l=n-1$  where $m$ is the magnetic quantum number, $l$ is the orbital quantum number, and $n$ is the radial quantum number. 
In a cryogenic environment the circular states have radiative lifetimes $\tau\sim n^5$ compared to $n^3$ for low angular momentum states. 
 The dipole-dipole interaction and the blockade shift for the high orbital angular momentum state is comparable with the low angular momentum state. Thus the intrinsic error for the quantum gate via Rydberg blockade will be suppressed. We present numerical simulations of quantum process tomography to find the intrinsic fidelity of a Rydberg blockade controlled phase gate using circular Rydberg states. Our analysis shows that the intrinsic gate error extracted from simulated quantum process tomography can be below $9 \times10^{-6}$ for specific states of Cs atoms in a cryogenic  environment. 

In Sec. \ref{sec.scheme} we present the scheme of a two-qubit quantum gate using circular Rydberg states. in Sec. \ref{sec.calculation}  we calculate the  dipole-dipole interaction between two alkali metal atoms in circular states as well as their lifetimes. In Sec. \ref{sec.errors} we give analytical estimates of the intrinsic gate error in the computational basis using circular states. 
In Sec. \ref{sec.simulation} we perform simulated quantum process tomography of a two qubit controlled-phase gate accounting only for intrinsic errors from Sec. \ref{sec.errors}. This analysis shows that in a well designed experiment where technical errors are minimized it should be possible to reach low gate errors, below fault tolerance thresholds. A discussion and summary is presented in Sec. \ref{sec.discussion}.

\section{A two qubit quantum gate with circular Rydberg states}
\label{sec.scheme}

\begin{figure}[!t]
\begin{centering}
\includegraphics[width=.9\columnwidth]{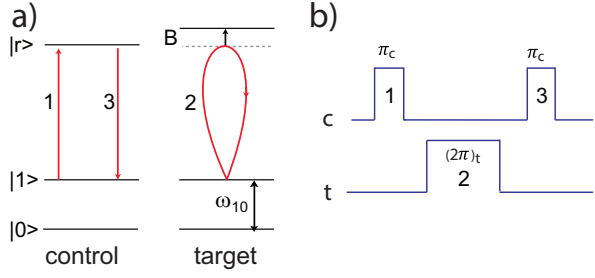}
\par\end{centering}
\caption{(Color online) (a) Scheme of implementing a two qubit quantum gate with circular Rydberg states. (b) Pulse sequence for the C$_Z$ gate (pulses 1-3).} 
\label{fig:scheme}
\end{figure}

The  scheme to implement a two qubit quantum gate is the same as in the original proposal by Jaksch, et al.  \cite{Jaksch2000} except that we use circular Rydberg states. Consider  two atoms (one is control and the other is target) separated by several $\mu\rm m$. We encode qubits in two internal atomic ground states (e.g. hyperfine states) denoted by $|1\rangle$ and $|0\rangle$  and    $|r\rangle$ is a circular Rydberg state as shown in Fig. \ref{fig:scheme}

%

A two qubit C$_Z$ gate is implemented  using a three pulse sequence between $|1\rangle$ and $|r\rangle$ (pulses 1-3 in Fig. \ref{fig:scheme} ): first we apply a Rydberg $\pi$ pulse  to the control atom, then a Rydberg $2\pi$ pulse to the target atom, and finally a Rydberg $\pi$ pulse to bring the control atom back to the $|1\rangle$ state. The strong interaction between atoms in $|r\rangle$ states gives a blockade shift $\sf B$ which blocks excitation of the target atom if the control atom has been Rydberg excited.  This leads to a conditional phase shift of the two-atom state which can be used to generate entanglement.

Although the pulse sequence is the same as has been demonstrated in experiments with low-angular momentum states\cite{Saffman2010} it is an outstanding technical challenge to rapidly excite the atoms to circular Rydberg states with high angular momenta. We will first calculate the achievable gate error assuming that we can coherently drive atoms between  $|1\rangle$ and Rydberg state $|r\rangle$ with high fidelity. 
We will return to the question of Rydberg excitation fidelity in Sec. \ref{sec.discussion}.

\section{Rydberg blockade shift and  lifetimes of circular states}
\label{sec.calculation}

The fidelity error of a Rydberg blockade quantum gate scales as\cite{Saffman2005a,Saffman2010} 
$ 1/({\sf B} \tau)^{2/3}$.  We therefore need to calculate the blockade shift $\sf B$ and lifetime $\tau$ for circular states. 
The Rydberg blockade effect arises from the dipole-dipole interaction between two atoms in Rydberg states. When one of the atoms is excited to a Rydberg state, the Rydberg level of the other atom is  shifted by the dipole-dipole interaction which blocks any subsequent excitation.

We write two-atom Rydberg states of atoms $A,B$ as $|n,l,m\rangle_A |n',l',m'\rangle_B$ and the circular state with radial quantum number $n$ as
$|c_n\rangle=|n,n-1,n-1\rangle$ .
When the atomic angular momentum is quantized in a coordinate system parallel to the molecular axis joining the atoms  
the dipole-dipole interaction preserves the angular momentum projection $m+m'$ and for 
 the symmetric circular state $|C\rangle=|c_nc_n\rangle=|c_n\rangle_A |c_n\rangle_B$, the nearest energy  state is $|C'\rangle=|c_{n+1}c_{n-1}\rangle=|c_{n+1}\rangle_A |c_{n-1}\rangle_B$ with the energy defect $\hbar\delta=\frac{E_{\rm H}}{2}\left(-\frac{1}{(n+1)^2}-\frac{1}{(n-1)^2}+\frac{2}{n^2}\right)$ where $E_{\rm H}$ is the Hartree energy.  The second nearest  state which is dipole coupled to $|C\rangle$ is $|n+2,n,n\rangle_A |c_{n-1}\rangle_B$ with the energy defect $\hbar\delta'=\frac{E_{\rm H}}{2}\left(-\frac{1}{(n+2)^2}-\frac{1}{(n-1)^2}+\frac{2}{n^2}\right)$. For $n=100$, $\delta=-3\times 10^{-8} E_{\rm H}$ while $\delta'=0.93\times 10^{-6} E_H$. Since  $\delta\sim  n^{-4}$ while $\delta'\sim  n^{-3}$, the ratio $\delta/\delta'$ tends to zero for large $n$. It is thus a good approximation to keep only the two states $|C\rangle$ and  $|C'\rangle$ in the  Rydberg blockade analysis.

With this approximation the Hamiltonian for the two level system $|C\rangle$ and $|C'\rangle$ is 
\begin{equation}
H=\begin{bmatrix}
0 & \hat V_{dd}^{(0)}\\
\hat V_{dd}^{(0)} & \delta
\end{bmatrix}.
\label{eq:ham}
\end{equation}
where the dipole-dipole interaction operator is 
$\hat V_{\rm dd}^{(0)} = -\frac{\sqrt6 e^2}{4\pi\epsilon_0 R^3}\sum_{p}C_{1p1-p}^{20}r_{Ap}r_{B-p}$ with 
matrix element 
\begin{eqnarray}
V_{dd}
&=& \langle c_{n+1}c_{n-1}|\hat{V}_{\rm dd}^{(0)}|c_nc_n\rangle
\nonumber\\
&=& \frac{-\sqrt{6}e^2}{4\pi \epsilon_0 R^3}
\frac{ \langle n+1n||r||nn-1\rangle \langle n-1n-2||r||nn-1\rangle}{\sqrt{(2n+1)(2n-3)}} \nonumber\\
&&\times ~  C^{20}_{111-1} C^{nn}_{n-1n-111} C^{n-2n-2}_{n-1n-11-1}
\nonumber\\
&=& \frac{e^2a_0^2}{4\pi \epsilon_0 R^3}\frac{8\,  2^{4n}n^{2n+4}(n^2-1)^{n+2}}{(2n+1)^{2n+3}(2n-1)^{2n+1}} .
\label{eq:Vdd}
\end{eqnarray}

Here $R$ is the separation between the atoms, $e$ is the elementary charge,  $a_0$ is the Bohr radius, and $C_{....}^{..}$ is a Clebsch-Gordan coefficient. The radial matrix elements were calculated using hydrogenic wavefunctions for which 
\begin{eqnarray}
\langle c_{n-1}||r||c_n\rangle&=&-\frac{4^{n} n^{n+1}(n-1)^{n+3/2}\sqrt{4n^2-6n+2}}{(2n-1)^{2n+1}}a_0\nonumber\\
\langle c_{n+1}||r||c_n\rangle&=& \frac{2^{1/2}4^{n+1} (n+1)^{n+2}n^{n+3}}{(2n+1)^{2n+5/2}}a_0\nonumber
\label{eq:radial}
\end{eqnarray}
For large $n$ we find the expected $n^4$ dipole-dipole scaling  $V_{\rm dd}\simeq \frac{e^2 a_0^2}{4\pi\epsilon_0 R^3 }8 n^4.$

\begin{figure}[!t]
\begin{centering}
\includegraphics[width=1\columnwidth]{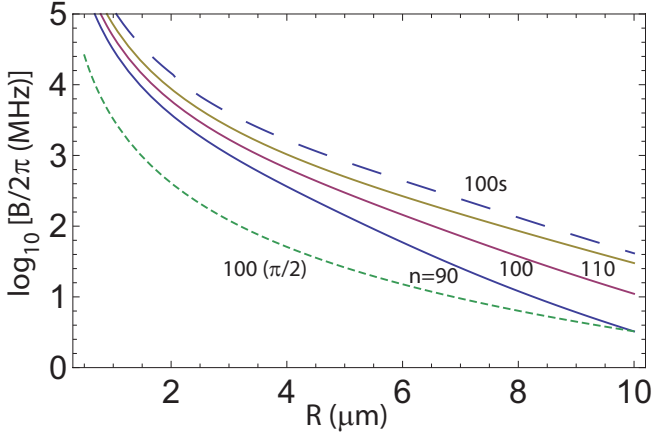}
\par\end{centering}
\caption{Blockade shift versus $R$ for circular states $n=90,100,110$ (solid lines), the $n=100$ circular state in a 90 deg. geometry (dotted line), and the  Cs $100s$ state (dashed line).  } 
\label{fig:B}
\end{figure}

The eigenvalues of the Hamiltonian (\ref{eq:ham}) are $ U_\pm=\frac{1}{2}\left(\delta\pm \sqrt{\delta^2+4V_{\rm dd}^2} \right).$ At large $R$ a pair of noninteracting atoms has zero energy so the effective blockade shift in the limit of negligible two-atom excitation, which is relevant for gate operation, is simply ${\sf B}=U_+$ (we take the plus sign since $\delta<0$).  The blockade shift is plotted in Fig. \ref{fig:B}
together with the blockade for a pair of atoms in low angular momentum states. We see that for the same principal quantum number the circular states have a much smaller blockade shift. As we will show below they are nonetheless useful for gate operations due to their much longer radiative lifetimes. 

When the quantization axis is perpendicular to the molecular axis the dipole-dipole operator is 
$$
\hat V_{\rm dd}^{(\pi/2)}  = -\frac{1}{2}\hat V_{\rm dd}^{(0)} -\frac{ e^2}{4\pi\epsilon_0 R^3}\frac{3}{2}\sum_{p}\left( C_{1p1p}^{2, 2p}-C_{1p1p}^{20}\right)r_{Ap}r_{Bp}.
$$
The selection rules are now $m+m'=0,\pm 2$ and there is a resonant interaction
 $\ket{c_nc_n}\leftrightarrow\ket{n,n-2,n-2}_A\ket{n,n-2,n-2}_B$. The matrix element is 
\begin{eqnarray}
V_{dd}&=& _A\langle n,n-2,n-2|_B\langle n,n-2,n-2|\hat{V}_{\rm dd}^{(\pi/2)}|c_nc_n\rangle\nonumber\\
&=&  -\frac{ e^2}{4\pi\epsilon_0 R^3}\frac{3}{2}
\frac{ \langle nn-2||r||nn-1\rangle^2}{2n-3} \nonumber\\
&&\times ~
 C_{1-11-1}^{2 -2} \left(C_{n-1n-11-1}^{n-2 n-2}\right)^2  
\nonumber\\
&=& \frac{e^2a_0^2}{4\pi \epsilon_0 R^3}\frac{27}{8}n^2(n-1),
\label{eq:Vdd90}
\end{eqnarray}
where we have used $\langle nn-2||r||nn-1\rangle=\frac{3}{2}n\sqrt{2n-1}\sqrt{n-1}$. For this geometry we get a much weaker  interaction scaling as $n^3$. Since it is resonant ${\sf B}=V_{\rm dd}$ and the interaction strength falls off as $1/R^3$ which is advantageous at long range. However we will be interested in values of $R<5~\mu\rm m$ and will therefore only consider the parallel geometry in the following.

\begin{figure}[!t]
\begin{centering}
\includegraphics[width=1\columnwidth]{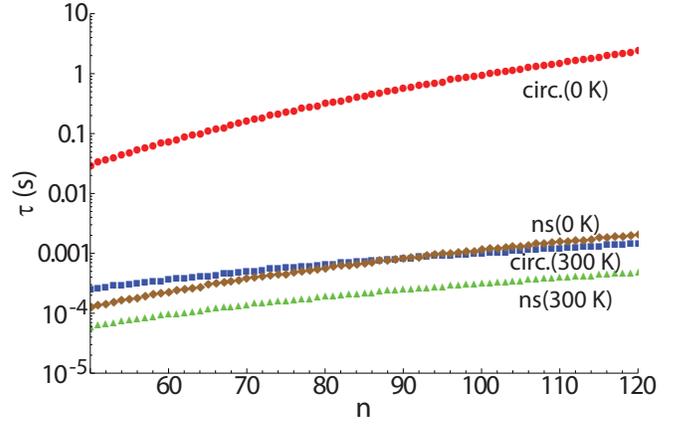}
\par\end{centering}
\caption{(color online) Radiative lifetime of the circular states and $ns$ states at 0 and 300 K. The $ns$ state lifetimes were calculated using approximate expressions given in Ref. \cite{Beterov2011}. } 
\label{fig:lt}
\end{figure}

The radiative lifetime for the circular state $|c_n\rangle$ due to 
  decay to the next circular state $|c_{n-1}\rangle$ is
\begin{equation}
\tau_0=\frac{3\pi \epsilon_0 \hbar c^3}{\omega_{eg}^3 e^2 |\langle c_{n-1}|r_{-1}|c_n\rangle|^2},
\label{eq:lt}
\end{equation}
where the transition frequency is $\omega_{eg}=\frac{E_R}{\hbar}\left[1/(n-1)^2-1/n^2\right]$. Using the expressions given above for the reduced matrix elements we find   
\begin{equation}
\tau_0=\frac{3\pi\epsilon_0 \hbar^4 c^3  }{ E_R^3 a_0^2 e^2}\frac{(2n-1)^{4n-1} }{2^{4n+1} n^{2n-4} (n-1)^{2n-2}}.
\label{eq:ltanalyt}
\end{equation}
This is the lifetime at zero temperature. The finite temperature blackbody correction gives  
\begin{equation}
\frac{1}{\tau}=\frac{1}{\tau_0}\left(\frac{1}{e^{\hbar \omega_{eg}/k_B T}-1}+1\right)
\label{eq:temperature}
\end{equation}
where $k_B$ is the Boltzmann constant and $T$ is the temperature. 
The circular state lifetimes from (\ref{eq:temperature}) are compared with the lifetime of Cs $ns$ states in Fig. \ref{fig:lt}.
Because the transition frequency $\omega_{eg}$ is in the microwave regime, the finite temperature correction factor is bigger than that for low angular momentum Rydberg states. 

%

\section{Intrinsic error estimates}
\label{sec.errors}

The intrinsic error of a Rydberg blockade C$_Z$ gate arises from decoherence due to 
the finite lifetime $\tau$ of the Rydberg state and state rotation errors due to imperfect blockade. 
In the strong blockade limit $\Omega\ll {\sf B}\ll\omega_{10}$ where $\Omega$ is the Rydberg state excitation frequency  
the intrinsic gate error $E_1$ averaged over the input states in the computational basis 
$(\ket{00},\ket{01},\ket{10},\ket{11})$ for the scheme shown in Fig. \ref{fig:scheme} is \cite{Saffman2005a, Saffman2010}
\begin{equation}
E_1=\frac{7\pi}{4\Omega\tau}\left(1+\frac{\Omega^{2}}{\omega_{10}^{2}}+\frac{\Omega^{2}}{{7\sf B}^{2}}\right)+\frac{\Omega^{2}}{8{\sf B}^{2}}\left(1+6\frac{{\sf B}^{2}}{\omega_{10}^{2}}\right)
\label{eq:IntrinsicError}
\end{equation}
The first term in Eq. (\ref{eq:IntrinsicError}) is the Rydberg decay error 
due to the finite lifetime $\tau$ of the Rydberg circular state,  and the second term is the imperfect blockade error. 
In the limit of $\omega_{10}\gg({\sf B},\Omega)$ 
we can extract a simple expression for the optimum Rabi frequency which minimizes the error
\begin{equation}
\Omega_{\rm opt}=(7\pi)^{1/3}\frac{{\sf B}^{2/3}}{\tau^{1/3}}.
\label{eq:OptRabi}
\end{equation}
Setting $\Omega\rightarrow\Omega_{\rm opt}$ leads to a minimum averaged gate error of
\begin{equation}
E_{\rm min}=\frac{3(7\pi)^{2/3}}{8}\frac{1}{({\sf B}\tau)^{2/3}}.
\label{eq:MinError}
\end{equation}
%

\begin{figure}[!t]
\begin{centering}
\includegraphics[width=1\columnwidth]{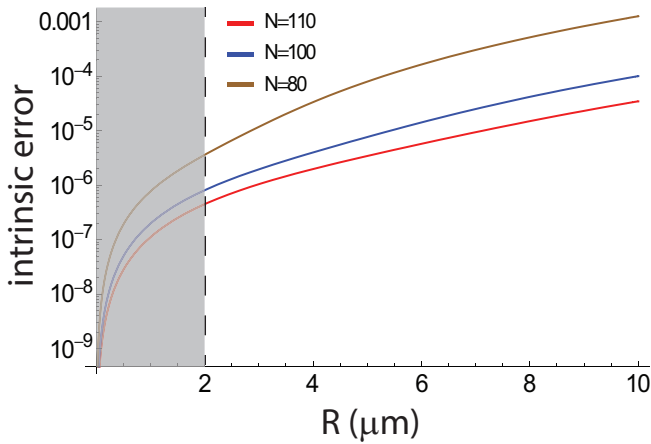}
\par\end{centering}
\caption{The minimum intrinsic error from Eq. (\ref{eq:MinError}) for $n=80,~ 100$, and $110$  as a function of the separation between the two atoms. The gray region is excluded due to Rydberg wavefunction overlap (see text).}
\label{fig:intrinserr}
\end{figure}

\begin{figure}[!t]
\begin{centering}
\includegraphics[width=.87\columnwidth]{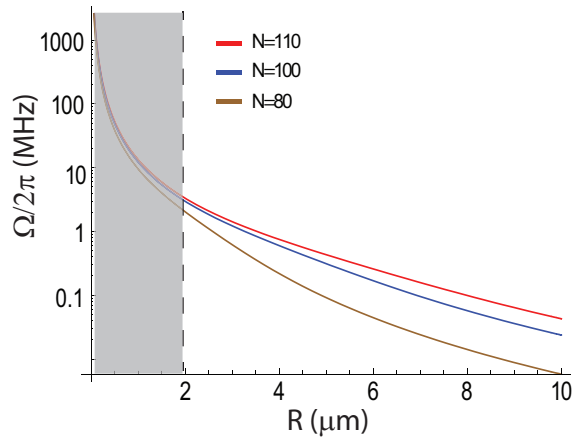}
\par\end{centering}
\caption{Optimal Rabi frequency from Eq. (\ref{eq:OptRabi}) for $n=80,~ 100$, and $110$  as a function of the separation between the two atoms. The gray region is excluded due to Rydberg wavefunction overlap (see text).}
\label{fig:omg}
\end{figure}

Figures \ref{fig:intrinserr}, \ref{fig:omg} show the calculated $E_{\rm min}$ and $\Omega_{\rm opt}$ for several states as a function of atomic separation $R$. Although the intrinsic error appears to become arbitrarily small at small $R$ we must impose a minimum value of $R$ to avoid overlap of the spatially extended Rydberg wavefunctions.  
The atomic size scales as $n^2$ and for $n=110$ and $l=109$, the peak of the radial wavefunction is at $0.64~ \mu \rm m$. The probability of finding the electron outside a sphere with radius $1~ \mu \rm m$ is less than $10^{-12}$. The electron overlap is thus negligible if  two Rydberg atoms are separated by $R=2~ \mu m$. With this condition,  the minimum intrinsic gate error in Eq. (\ref{eq:MinError}) is $1.6\times 10^{-7}$ for $n=110$. Compared with low angular momentum states $112p_{3/2}$ and $112d_{5/2}$ \cite{XZhang2012}, the circular Rydberg states improve the minimum intrinsic error by about three orders of magnitude.

\section{Simulated Quantum Process Tomography}
\label{sec.simulation}

While the intrinsic error estimates presented above provide some guidance, the performance of a quantum gate is also dependent on phase errors which are not captured by the intrinsic error  estimate. Full process tomography simulations show that entangling gate fidelities may be more than an order of magnitude larger than the above estimates \cite{XZhang2012}. We therefore present process tomography simulations in order to determine the achievable gate performance. 
In this analysis we only account for intrinsic gate errors as described in Sec. \ref{sec.errors},  and assume all additional technical errors are negligible. This corresponds to a situation where the atoms are cooled to their motional ground state and are held in magic traps for both the ground and Rydberg states\cite{SZhang2011,Morrison2012} so there is no Doppler dephasing during Rydberg excitation,  position dependent variations in Rabi frequencies, or AC Stark shifts. We also  assume that we can coherently transfer atoms between $|1\rangle$ and $|r\rangle$ states, and  that dephasing due to time varying magnetic fields is negligible.

A reliable method to characterize the performance of quantum gates is Quantum Process Tomography (QPT) \cite{Chuang1997,*Poyatos1997,Nielsen2000}. QPT has been demonstrated with several different physical systems 
including linear optics \cite{OBrien2004,*White2007}, trapped ions \cite{Riebe2006,*SXWang2010}, and superconducting circuits \cite{Yamamoto2010,*Bialczak2010}. It was also used to numerically simulate the performance of a Rydberg-blockade $C_Z$ gate using low angular momentum Rydberg states in Ref. \cite{Zhang2012}. Here, we follow the same procedures as in Ref. \cite{Zhang2012}, but for circular Rydberg states.

We use Cesium in the numerical calculations and for each atom we include four atomic states (see Fig. \ref {fig:scheme}): qubit $|0\rangle$, qubit $|1\rangle$,  and reservoir level $|g\rangle\equiv|{6s}_{1/2},m_{F}\neq0\rangle$ in the ${6s}_{1/2}$ ground state, and the Rydberg circular state $|r\rangle$. With this set of basis states the two-atom dynamics are described by density matrices $\rho_{\rm ct}(t)$ with dimensions $16\times 16$. We take the initial condition to be a separable state $\rho_{\rm ct}(0)=\rho_{\rm c}(0)\otimes\rho_{\rm t}(0)$, with c/t for control/target atoms.
We calculate the time evolution by solving the master equation
\begin{equation}
\frac{d\rho_{\rm ct}}{dt}=-\frac{i}{\hbar}[H_{\rm ct},\rho_{\rm ct}]+\mathcal{L_{\textrm{ct}}},
\label{eq:master}
\end{equation}
with ${H}_{\rm ct}={H}_{\rm c}\otimes I_{\rm t}+I_{\rm c}\otimes{H}_{\rm t}+\hbar{\sf B}\left[\begin{array}{cc}
0_{15} & 0\\0 & 1\end{array}\right]$, 
$\mathcal{L}_{\rm ct}=\mathcal{L}_{\rm c}\otimes I_{\rm t}+I_{\rm c}\otimes\mathcal{L}_{\rm t}$, $I_{\rm t}$ $(I_{\rm c})$ 
are  $4\times4$ identity matrices, and 0$_{15}$ is a  $15\times15$ zero matrix.  After making the rotating-wave approximation the  Hamiltonian ${H}_{\rm c}$ (${H}_{\rm t}$),
and the Liouville operators  $\mathcal{L_{\textrm{c}}}$ ($\mathcal{L_{\textrm{t}}}$)
are given in the basis $\{|0\rangle,|g\rangle,|1\rangle,|r\rangle$\} as
%
\begin{subequations}
\begin{eqnarray}
 H_{\rm (c/t)}  & = & \hbar \left(\begin{array}{cccccc}
-\omega_{10} & 0  & 0 & \Omega_{\rm (c/t)}^{*}/2 \\
0 & 0 & 0 & 0  \\
0 & 0 & 0  & \Omega_{\rm (c/t)}^{*}/2 \\
\Omega_{\rm (c/t)}/2 & 0   &  \Omega_{\rm (c/t)}/2 & 0 \end{array}\right),
\label{eq:Hamiltonian2}\\
\mathcal{L}_{\rm (c/t)}&=&\gamma_r\left(\begin{array}{cccccc}
\frac{1}{16}\rho_{rr} & 0 & 0  & -\frac{1}{2}\rho_{0r}   \\
0 & \frac{7}{8}\rho_{rr} & 0 & -\frac{1}{2} \rho_{gr}  \\
0  & 0 & \frac{1}{16}\rho_{rr} & -\frac{1}{2} \rho_{1r}   \\
-\frac{1}{2} \rho_{r0}  & -\frac{1}{2} \rho_{rg}  & -\frac{1}{2} \rho_{r 1}  & -\rho_{rr} 
\end{array}\right).
\label{eq:Liouville2}
\end{eqnarray}
\label{eq:HL2}
\end{subequations}
We assume that the Rydberg states decay directly back to the 16 ground sublevels of Cs with equal branching ratios of $1/16$.  

The details of simulated QPT of the $C_Z$ gate can be found in Ref. \cite{Zhang2012}.  Here we give a brief overview of the procedure.  We start with 16 linearly independent input states with both atoms in one of the four states ($|0\rangle$, $|1\rangle$, ($|0\rangle$+$|1\rangle$)/$\sqrt{2}$, and ($|0\rangle$+$\imath |1\rangle$)/$\sqrt{2}$. 
We then solve the time evolution of the master equation (\ref{eq:master}) for the $C_Z$ pulse sequence of Fig. \ref{fig:scheme}b for each of the 
input states
 $\{\pi_{\rm c},(2\pi)_{\rm t},\pi_{\rm c}\}$, where $(\pi)_{\rm {c}}$ is a $\pi$ pulse between $|1\rangle$ and $|r\rangle$ for the control atom, and $(2\pi)_{\rm {t}}$ is a 2$\pi$ pulse between $|1\rangle$ and $|r\rangle$ for the target atom. The output states found in this way may be 
non-physical.  We then perform  maximum likelihood estimation (MLE) \cite{OBrien2004} to reconstruct physical states. This process is so-called Quantum State Tomography (QST). From the QST,  we can extract a physical $\chi$-matrix for the simulated $C_Z$ gate using a maximum likelihood estimator \cite{OBrien2004, Nielsen2002,*Pedersen2007}. Finally we  quantify the performance of the simulated $C_Z$ gate from the $\chi$-matrix.

 A widely used measure of quantum processes is the trace overlap fidelity $F_{\rm O}$, or error $E_{\rm O}=1-F_{\rm O}$ 
which are based on the trace overlap between ideal and experimental (in our case simulated) $\chi$ process matrices. 
The fidelity error is defined by
\begin{equation}
E_{\rm O}  = 1- {\rm Tr}^2\left[\sqrt{\sqrt{\chi_{\rm sim}}\chi_{\rm id}\sqrt{\chi_{\rm sim}}}\right],
\label{eq:error2}
\end{equation}
where $\chi_{\rm id}$ is the ideal process matrix and $\chi_{\rm sim}$ is the simulated physical $\chi$-matrix found from QPT accounting for  intrinsic gate errors as described by Eqs. (\ref{eq:HL2}).

\begin{table*}[!t]
\caption{Gate errors from simulated QPT for several circular Rydberg states of  $^{133}$Cs. The reported errors are $E_{\rm cb}$, the analytical estimate found in Sec. \ref{sec.errors} using computational basis states, trace loss, which is the sum of populations outside the computational basis at the end of the gate sequence, and $E_{\rm O}$ trace overlap errors from Eq. (\ref{eq:error2}). }

\begin{tabular}{|l|l|l|llll|llll|}
\hline
& $^{133}$Cs (n=80) &$^{133}$Cs (n=100) &$^{133}$Cs (n=110)&&&\\
\hline 
Temperature (K) &$0$ &$0$&$0$&$77$ &$300$ &\tabularnewline
Rabi frequency $\Omega/2\pi~\rm (MHz)$& 3.82& 5.05& 5.6 &38.4&60.3 & \tabularnewline
Blockade shift  ${\sf B}/2\pi~\rm (GHz)$& 2.21& 5.89& 8.71 &8.71&8.71 & \tabularnewline
Lifetime  (ms)&307&940& 1520 &4.71&1.21 & \tabularnewline
Trap separation $ (\mu \rm m)$& 2& 2& 2 &2&2 & \tabularnewline
 $E_{{\rm  cb}}$   & $1.1\times 10^{-6}$& $2.8\times 10^{-7}$& $1.6\times 10^{-7}$ &$7.3\times 10^{-6}$&$1.8\times 10^{-5}$&  \tabularnewline
 trace loss   & $5.1\times 10^{-6}$& $1.3\times 10^{-6}$& $7.0\times 10^{-7}$ &$3.3\times 10^{-5}$&$8.1\times 10^{-5}$& \tabularnewline
 $E_{{\rm  O}}$   & $2.6\times 10^{-5}$& $1.9\times 10^{-5}$& $8.8\times 10^{-6}$ & $1.1\times 10^{-4}$&$2.3\times 10^{-4}$&   \tabularnewline

\hline
\end{tabular}
\label{tab:QPTerrors}
\end{table*}

In Table \ref{tab:QPTerrors} we present the errors found from simulated QPT for the listed atomic states. The process tomography errors tend to be one to two orders of magnitude larger than $E_{\rm cb}$ which are the errors estimated in Sec. \ref{sec.errors} for two-qubit product states in the  computational basis. This is to be expected since the analytical estimates are derived from the probabilities of the gate succeeding, and do not account for output state phase errors.  The trace loss quantifies the population in states outside the computational basis at the end of the gate sequence. These errors are due to spontaneous emission from Rydberg states and imperfect blockade which leaves atoms Rydberg excited at the end of the gate. The process error based on trace overlap $E_{\rm O}$ can be as low as $3.4\times 10^{-5}$ for the $n=110$ circular Rydberg state in a 
cryogenic environment.     

\section{Discussion}
\label{sec.discussion}

We have proposed and simulated a Rydberg blockade mediated two qubit quantum gate between two individually addressed neutral atoms using circular Rydberg states. We show that the gate error based on simulated QPT for Cs atom states can be at the level of $4.3\times 10^{-4}$ for the $n=110$ circular Rydberg state at room temperature.  With the help of a cryostat the process error can be as low as $3.4\times 10^{-5}$. These small error numbers can be contrasted with the optimal result found for low angular momentum Rydberg states in \cite{XZhang2012} which was $\sim 1\times 10^{-3}$. 
While the use of circular states can potentially reduce the gate error by more than a factor of 100 the circular states present challenges for practical use. We discuss these issues in the following sections. 

\subsection{Excitation of circular states}

Of course in order to implement such a gate it is necessary to coherently excite circular Rydberg states on a fast time scale with very high fidelity. 
 The production of circular states 
has been demonstrated with Lithium \cite{Hulet1983, Hare1988, Nussenzweig1991}, Rubidium \cite{Brecha1993,Nussenzweig1993}, and Sodium \cite{Cheng1994} atoms. Of particular relevance to the ideas proposed here cold Rb atoms have been recently excited to Rydberg states and magnetically trapped \cite{Anderson2013}. There are  two main methods to produce circular states both of which start from low angular momentum Rydberg states. The first method is called microwave adiabatic transfer\cite{Molander1986, Hulet1983, Nussenzweig1991, Nussenzweig1993, Lutwak1997}. It holds the microwave frequency constant and makes the frequency resonant with transitions from $m$ to $m+1$ by way of the  second order Stark effect in a time varying electric field. Under the constant microwave frequency and the varying amplitude  electric field, the atoms are transferred to the circular states via a series of adiabatic passages. The second method makes use of crossed electric and magnetic fields\cite{Delande1988, Hare1988, Brecha1993, Lutwak1997}. The atoms start from an $m=0$ state with a  large electric field and weak magnetic field. When the electric field is gradually decreased to zero and the crossed magnetic field is constant, the atom is adiabatically transferred from the largest electric dipole energy to the largest magnetic dipole energy, which just corresponds to a circular state with maximal $m$. The efficiency of circular state transfer can be nearly $100\%$ \cite{Hulet1983, Lutwak1997}.

The microwave adiabatic transfer and crossed electromagnetic field approaches can in principle be rapid. Nevertheless they are not appropriate for  gate operation since they result in substantial population of intermediate states. In such a case Rydberg blockade of the final state will result in population being left behind in a Rydberg excited state leading to a large gate error. Stimulated Raman adiabatic passage (STIRAP) provides a promising alternative approach. STIRAP h	as been widely applied to three level atoms for  transfer from an  initial state to a final state without populating the intermediate state. As has been shown by Vitanov the idea of dark state transfer can be generalized to multistate problems \cite{Vitanov1998}.   In order to prevent population of the intermediate states all excitation fields must be detuned from all intermediate states. In \cite{Vitanov1998}  there is a detailed discussion of the off-resonant case using a counterintuitive pulse sequence. The Stokes pulse precedes the pump pulse, where the Stokes pulse $\Omega_S$ couples the final state $|\psi_N\rangle$ and the last intermediate state $|\psi_{N-1}\rangle$ and the pump pulse $\Omega_P$ couples the initial state $|\psi_1\rangle$ and the first intermediate state $|\psi_2\rangle$. These two pulses are shaped, so that the pump pulse has a time delay but still has an overlap with the Stokes pulse.  Different pulse sequences  for the intermediate pulses $\Omega_{k,k+1}$ which couple the neighboring intermediate states $|\psi_k\rangle$ and $|\psi_{k+1}\rangle$ are possible as discussed in \cite{Vitanov1998}.

\begin{figure}[!t]
\begin{centering}
\includegraphics[width=1\columnwidth]{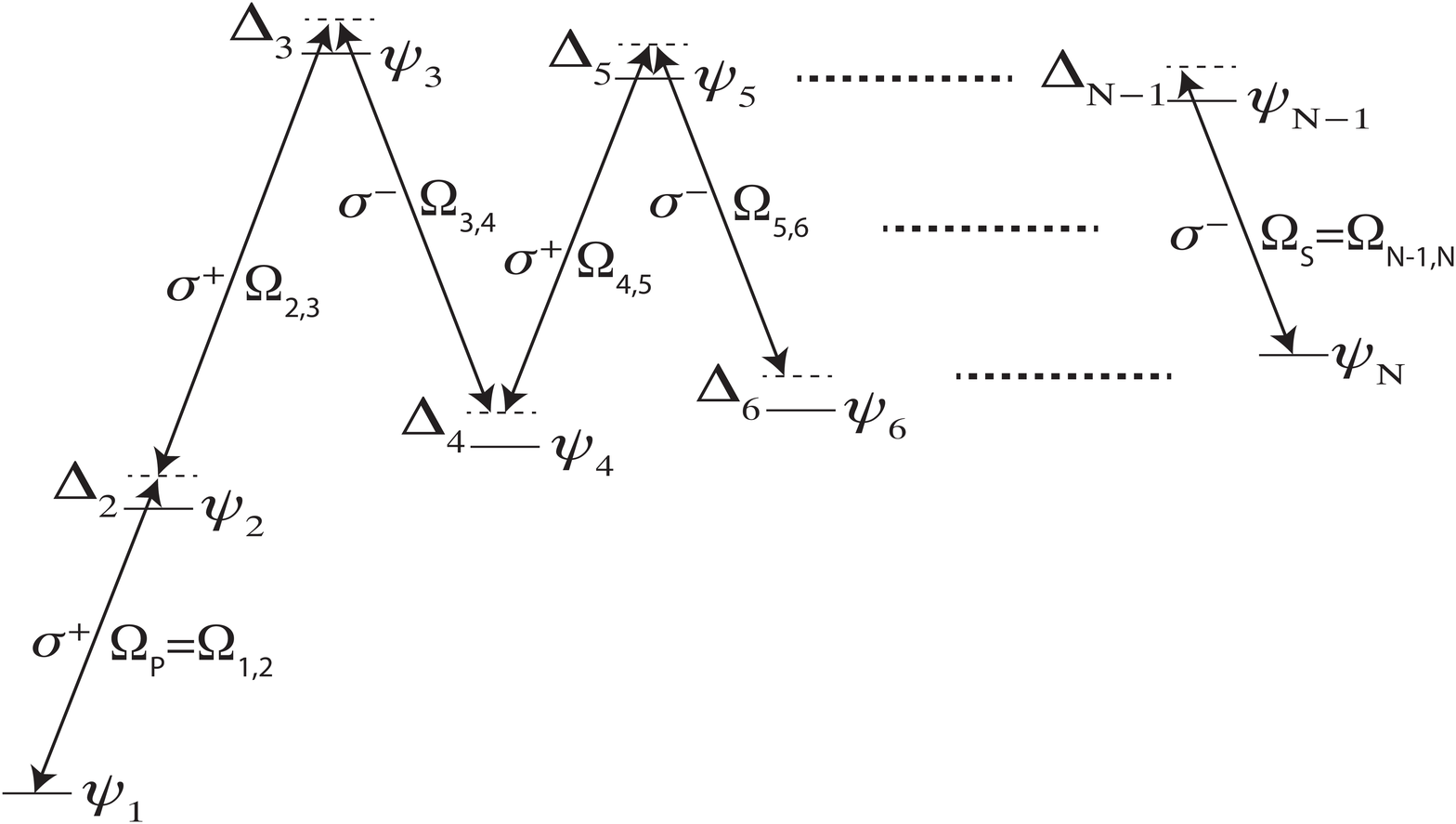}
\par\end{centering}
\caption{Illustration of multiphoton STIRAP process for  transfer from the ground state $|\psi_1\rangle$ to a circular Rydberg state $|\psi_N\rangle$ in Cs. The intermediate states are  $|\psi_{2}\rangle=|7p_{1/2}, F=4, m_F=1\rangle$ and two chains of Rydberg states. The odd numbered chain consists of the states  $|\psi_{2k-1}\rangle=|n=170-k, l=2k-2, m=2k-2\rangle$ starting from $|\psi_{3}\rangle=|168, 2, 2\rangle$ $(k=2)$,  and ending  with $|\psi_{N-1}\rangle=|\psi_{111}\rangle=|114, 110, 110\rangle$   
 $(k=56)$. The even numbered  chain consists of the states  $|\psi_{2k}\rangle=|n=56+k, l=2k-1, m=2k-1\rangle$ starting from $|\psi_{4}\rangle=|58, 3, 3\rangle$, $(k=2)$, 
and ending with the final state $|\psi_{N}\rangle=|\psi_{112}\rangle=|112, 111, 111\rangle$  $(k=56)$. The frequencies needed for the STIRAP chain coupling $\psi_3\rangle$ to $\psi_112$ range from 859 to 9.1 GHz.}
\label{fig:stirap}
\end{figure}

A possible implementation for exciting $|c_{112}\rangle$ is shown in Fig. \ref{fig:stirap}. Let us assume all intermediate pulses are  constant in  time. We use the qubit state 
$|1\rangle = |6s_{1/2}, F=4, m_F=0\rangle$ as the initial state 
$|\psi_{1}\rangle$ and the circular state $|c_{112}\rangle$   as the final state $|\psi_{N=112}\rangle$.  With the choice of states in Fig. \ref{fig:stirap}, $\Omega_P=\Omega_{1,2}$ is an optical pulse at 459 nm with $\sigma^{+}$ polarization, $\Omega_{2,3}$ is an optical pulse at 1038 nm with $\sigma^{+}$ polarization, $\Omega_{2k,2k+1}$ is a microwave pulse with $\sigma^{+}$ polarization and $\Omega_{2k-1,2k}$ is a microwave pulse with $\sigma^{-}$ polarization when $k\geq 2$. All of the single photon microwave frequencies are non-degenerate, so all the single photon Rabi frequencies could be controlled independently. We choose the signal and pump Rabi frequencies ($\alpha$ in Eq. (6) of \cite{Vitanov1998}) to be 14 MHz, the intermediate Rabi frequencies $\xi_{k,k+1}=100~\rm MHz$ for all the $k$ in Eq. (7) of \cite{Vitanov1998} and 
the intermediate detunings to be $\Delta=90\times \alpha$ in Eq. (24a) of \cite{Vitanov1998}. With these  parameters, Eq. (41a) of \cite{Vitanov1998} shows that  the overall Rabi frequency for transfer from $|\psi_1\rangle$ to $|\psi_N\rangle$ could be as large as $\Omega=2\pi\times 5$ MHz, and Eq. (45) of \cite{Vitanov1998} shows that the population summed over all intermediate states can be suppressed to as low as $P_{\rm int}\sim 10^{-4}$. Since the intermediate states are high lying Rydberg levels with average lifetimes $\tau_{\rm int}> 100~\mu\rm s$ (see Fig. \ref{fig:lt}) we estimate the spontaneous emission error from the intermediate states in a $\pi$ pulse to be $\pi P_{\rm int}/(\Omega \tau_{\rm int})\sim \pi 10^{-4}/ (2\pi \times 5. \times  100.)\sim 10^{-7}$ which is small compared to the gate process error in Table \ref{tab:QPTerrors}. 

Note that the first pulse, which is optical, can be focused to selectively excite control or target qubits. All subsequent pulses are at microwave frequencies, and therefore give off-resonant AC Stark shifts to the qubits, but negligible population transfer out of the computational basis. These AC Stark shifts are in  principle known and if necessary can be compensated with additional off-resonant laser pulses. The very large electric dipole matrix elements between Rydberg states which scale as $n^4$ will mitigate power requirements for fast state transfer.

\subsection{Errors due to excitation of other Rydberg states}

Inspection of Table \ref{tab:QPTerrors} shows that the best gate performance is obtained when the blockade interaction is very large, about 8.7 GHz at $n=110$. As was pointed out in \cite{XZhang2012} blockade induced level shifts can lead to excitation of a neighboring, non-targeted Rydberg level leading to additional gate errors. Such errors were accounted for in \cite{XZhang2012} by extending the Hilbert used for simulation of QPT to include additional Rydberg levels. 

While a  similar procedure could be followed here we argue that it is not necessary for the multiphoton excitation process described in the preceding section. The frequency separation between $|c_{110}\rangle$ and  $|c_{109}\rangle$
is 5.0 GHz. This implies that if the control atom is excited to $|c_{110}\rangle$ by the first pulse of the $C_Z$ gate sequence then $|c_{110}\rangle$ will be off-resonance for the target atom by 8.7 GHz, but other states will be shifted up in energy to a position less than 8.7 GHz from $|c_{110}\rangle$ which would lead to a smaller effective blockade. 

This situation must be accounted for when analyzing low angular momentum states excited by a one or two photon transition. Here we use a multiphoton process to end up in a 
state with definite $l,m$. Any state at lower energy than a circular state will have $l'<l$ and $m'<m$ and therefore will not be populated due to angular momentum selection rules, even though the effective detuning of such states is reduced by the blockade interaction.

In conclusion we have analyzed the use of circular Rydberg states for implementing quantum gates using Rydberg blockade. 
The circular states have the potential of gate errors at the level of $10^{-5}$, a factor of  100  times lower  than what is possible with  low angular momentum states. This would put 
the Rydberg blockade gate deep in the regime of fault tolerant quantum computing architectures. The use of circular states entails significant experimental challenges related to the requirement of fast, and coherent excitation. While the required capabilities are not particularly close to what has been demonstrated to date,  with ongoing developments in laser cooling and trapping techniques and frequency agile laser and microwave sources experiments along the lines outlined here may become possible.

\begin{acknowledgments}
This work was supported by NSF award PHY-1104531, the AFOSR Quantum Memories
MURI, and the IARPA MQCO program through ARO contract W911NF-10-1-0347.

\end{acknowledgments}


\begin{thebibliography}{52}%
\makeatletter
\providecommand \@ifxundefined [1]{%
 \@ifx{#1\undefined}
}%
\providecommand \@ifnum [1]{%
 \ifnum #1\expandafter \@firstoftwo
 \else \expandafter \@secondoftwo
 \fi
}%
\providecommand \@ifx [1]{%
 \ifx #1\expandafter \@firstoftwo
 \else \expandafter \@secondoftwo
 \fi
}%
\providecommand \natexlab [1]{#1}%
\providecommand \enquote  [1]{``#1''}%
\providecommand \bibnamefont  [1]{#1}%
\providecommand \bibfnamefont [1]{#1}%
\providecommand \citenamefont [1]{#1}%
\providecommand \href@noop [0]{\@secondoftwo}%
\providecommand \href [0]{\begingroup \@sanitize@url \@href}%
\providecommand \@href[1]{\@@startlink{#1}\@@href}%
\providecommand \@@href[1]{\endgroup#1\@@endlink}%
\providecommand \@sanitize@url [0]{\catcode `\\12\catcode `\$12\catcode
  `\&12\catcode `\#12\catcode `\^12\catcode `\_12\catcode `\%12\relax}%
\providecommand \@@startlink[1]{}%
\providecommand \@@endlink[0]{}%
\providecommand \url  [0]{\begingroup\@sanitize@url \@url }%
\providecommand \@url [1]{\endgroup\@href {#1}{\urlprefix }}%
\providecommand \urlprefix  [0]{URL }%
\providecommand \Eprint [0]{\href }%
\providecommand \doibase [0]{http://dx.doi.org/}%
\providecommand \selectlanguage [0]{\@gobble}%
\providecommand \bibinfo  [0]{\@secondoftwo}%
\providecommand \bibfield  [0]{\@secondoftwo}%
\providecommand \translation [1]{[#1]}%
\providecommand \BibitemOpen [0]{}%
\providecommand \bibitemStop [0]{}%
\providecommand \bibitemNoStop [0]{.\EOS\space}%
\providecommand \EOS [0]{\spacefactor3000\relax}%
\providecommand \BibitemShut  [1]{\csname bibitem#1\endcsname}%
\let\auto@bib@innerbib\@empty
\bibitem [{\citenamefont {Jaksch}\ \emph {et~al.}(2000)\citenamefont {Jaksch},
  \citenamefont {Cirac}, \citenamefont {Zoller}, \citenamefont {Rolston},
  \citenamefont {C\^ot\'e},\ and\ \citenamefont {Lukin}}]{Jaksch2000}%
  \BibitemOpen
  \bibfield  {author} {\bibinfo {author} {\bibfnamefont {D.}~\bibnamefont
  {Jaksch}}, \bibinfo {author} {\bibfnamefont {J.~I.}\ \bibnamefont {Cirac}},
  \bibinfo {author} {\bibfnamefont {P.}~\bibnamefont {Zoller}}, \bibinfo
  {author} {\bibfnamefont {S.~L.}\ \bibnamefont {Rolston}}, \bibinfo {author}
  {\bibfnamefont {R.}~\bibnamefont {C\^ot\'e}}, \ and\ \bibinfo {author}
  {\bibfnamefont {M.~D.}\ \bibnamefont {Lukin}},\ }\href@noop {} {\bibfield
  {journal} {\bibinfo  {journal} {Phys. Rev. Lett.}\ }\textbf {\bibinfo
  {volume} {85}},\ \bibinfo {pages} {2208} (\bibinfo {year}
  {2000})}\BibitemShut {NoStop}%
\bibitem [{\citenamefont {Saffman}\ \emph {et~al.}(2010)\citenamefont
  {Saffman}, \citenamefont {Walker},\ and\ \citenamefont
  {M\o{}lmer}}]{Saffman2010}%
  \BibitemOpen
  \bibfield  {author} {\bibinfo {author} {\bibfnamefont {M.}~\bibnamefont
  {Saffman}}, \bibinfo {author} {\bibfnamefont {T.~G.}\ \bibnamefont {Walker}},
  \ and\ \bibinfo {author} {\bibfnamefont {K.}~\bibnamefont {M\o{}lmer}},\
  }\href@noop {} {\bibfield  {journal} {\bibinfo  {journal} {Rev. Mod. Phys.}\
  }\textbf {\bibinfo {volume} {82}},\ \bibinfo {pages} {2313} (\bibinfo {year}
  {2010})}\BibitemShut {NoStop}%
\bibitem [{\citenamefont {Lukin}\ \emph {et~al.}(2001)\citenamefont {Lukin},
  \citenamefont {Fleischhauer}, \citenamefont {Cote}, \citenamefont {Duan},
  \citenamefont {Jaksch}, \citenamefont {Cirac},\ and\ \citenamefont
  {Zoller}}]{Lukin2001}%
  \BibitemOpen
  \bibfield  {author} {\bibinfo {author} {\bibfnamefont {M.~D.}\ \bibnamefont
  {Lukin}}, \bibinfo {author} {\bibfnamefont {M.}~\bibnamefont {Fleischhauer}},
  \bibinfo {author} {\bibfnamefont {R.}~\bibnamefont {Cote}}, \bibinfo {author}
  {\bibfnamefont {L.~M.}\ \bibnamefont {Duan}}, \bibinfo {author}
  {\bibfnamefont {D.}~\bibnamefont {Jaksch}}, \bibinfo {author} {\bibfnamefont
  {J.~I.}\ \bibnamefont {Cirac}}, \ and\ \bibinfo {author} {\bibfnamefont
  {P.}~\bibnamefont {Zoller}},\ }\href@noop {} {\bibfield  {journal} {\bibinfo
  {journal} {Phys. Rev. Lett.}\ }\textbf {\bibinfo {volume} {87}},\ \bibinfo
  {pages} {037901} (\bibinfo {year} {2001})}\BibitemShut {NoStop}%
\bibitem [{\citenamefont {Saffman}\ and\ \citenamefont
  {Walker}(2005{\natexlab{a}})}]{Saffman2005b}%
  \BibitemOpen
  \bibfield  {author} {\bibinfo {author} {\bibfnamefont {M.}~\bibnamefont
  {Saffman}}\ and\ \bibinfo {author} {\bibfnamefont {T.~G.}\ \bibnamefont
  {Walker}},\ }\href@noop {} {\bibfield  {journal} {\bibinfo  {journal} {Phys.
  Rev. A}\ }\textbf {\bibinfo {volume} {72}},\ \bibinfo {pages} {042302}
  (\bibinfo {year} {2005}{\natexlab{a}})}\BibitemShut {NoStop}%
\bibitem [{\citenamefont {Isenhower}\ \emph {et~al.}(2011)\citenamefont
  {Isenhower}, \citenamefont {Saffman},\ and\ \citenamefont
  {M\o{}lmer}}]{Isenhower2011}%
  \BibitemOpen
  \bibfield  {author} {\bibinfo {author} {\bibfnamefont {L.}~\bibnamefont
  {Isenhower}}, \bibinfo {author} {\bibfnamefont {M.}~\bibnamefont {Saffman}},
  \ and\ \bibinfo {author} {\bibfnamefont {K.}~\bibnamefont {M\o{}lmer}},\
  }\href@noop {} {\bibfield  {journal} {\bibinfo  {journal} {Quant. Inf.
  Proc.}\ }\textbf {\bibinfo {volume} {10}},\ \bibinfo {pages} {755} (\bibinfo
  {year} {2011})}\BibitemShut {NoStop}%
\bibitem [{\citenamefont {Wu}\ \emph {et~al.}(2010)\citenamefont {Wu},
  \citenamefont {Yang},\ and\ \citenamefont {Zheng}}]{Wu2010}%
  \BibitemOpen
  \bibfield  {author} {\bibinfo {author} {\bibfnamefont {H.-Z.}\ \bibnamefont
  {Wu}}, \bibinfo {author} {\bibfnamefont {Z.-B.}\ \bibnamefont {Yang}}, \ and\
  \bibinfo {author} {\bibfnamefont {S.-B.}\ \bibnamefont {Zheng}},\ }\href@noop
  {} {\bibfield  {journal} {\bibinfo  {journal} {Phys. Rev. A}\ }\textbf
  {\bibinfo {volume} {82}},\ \bibinfo {pages} {034307} (\bibinfo {year}
  {2010})}\BibitemShut {NoStop}%
\bibitem [{\citenamefont {M\o{}ller}\ \emph {et~al.}(2008)\citenamefont
  {M\o{}ller}, \citenamefont {Madsen},\ and\ \citenamefont
  {M\o{}lmer}}]{Moller2008}%
  \BibitemOpen
  \bibfield  {author} {\bibinfo {author} {\bibfnamefont {D.}~\bibnamefont
  {M\o{}ller}}, \bibinfo {author} {\bibfnamefont {L.~B.}\ \bibnamefont
  {Madsen}}, \ and\ \bibinfo {author} {\bibfnamefont {K.}~\bibnamefont
  {M\o{}lmer}},\ }\href@noop {} {\bibfield  {journal} {\bibinfo  {journal}
  {Phys. Rev. Lett.}\ }\textbf {\bibinfo {volume} {100}},\ \bibinfo {pages}
  {170504} (\bibinfo {year} {2008})}\BibitemShut {NoStop}%
\bibitem [{\citenamefont {M\"uller}\ \emph {et~al.}(2009)\citenamefont
  {M\"uller}, \citenamefont {Lesanovsky}, \citenamefont {Weimer}, \citenamefont
  {B\"uchler},\ and\ \citenamefont {Zoller}}]{Muller2009}%
  \BibitemOpen
  \bibfield  {author} {\bibinfo {author} {\bibfnamefont {M.}~\bibnamefont
  {M\"uller}}, \bibinfo {author} {\bibfnamefont {I.}~\bibnamefont
  {Lesanovsky}}, \bibinfo {author} {\bibfnamefont {H.}~\bibnamefont {Weimer}},
  \bibinfo {author} {\bibfnamefont {H.~P.}\ \bibnamefont {B\"uchler}}, \ and\
  \bibinfo {author} {\bibfnamefont {P.}~\bibnamefont {Zoller}},\ }\href@noop {}
  {\bibfield  {journal} {\bibinfo  {journal} {Phys. Rev. Lett.}\ }\textbf
  {\bibinfo {volume} {102}},\ \bibinfo {pages} {170502} (\bibinfo {year}
  {2009})}\BibitemShut {NoStop}%
\bibitem [{\citenamefont {Saffman}\ and\ \citenamefont
  {M\o{}lmer}(2009)}]{Saffman2009b}%
  \BibitemOpen
  \bibfield  {author} {\bibinfo {author} {\bibfnamefont {M.}~\bibnamefont
  {Saffman}}\ and\ \bibinfo {author} {\bibfnamefont {K.}~\bibnamefont
  {M\o{}lmer}},\ }\href@noop {} {\bibfield  {journal} {\bibinfo  {journal}
  {Phys. Rev. Lett.}\ }\textbf {\bibinfo {volume} {102}},\ \bibinfo {pages}
  {240502} (\bibinfo {year} {2009})}\BibitemShut {NoStop}%
\bibitem [{\citenamefont {Chen}(2011)}]{AChen2011}%
  \BibitemOpen
  \bibfield  {author} {\bibinfo {author} {\bibfnamefont {A.}~\bibnamefont
  {Chen}},\ }\href@noop {} {\bibfield  {journal} {\bibinfo  {journal} {Opt.
  Express}\ }\textbf {\bibinfo {volume} {19}},\ \bibinfo {pages} {2037}
  (\bibinfo {year} {2011})}\BibitemShut {NoStop}%
\bibitem [{\citenamefont {M\o{}lmer}\ \emph {et~al.}(2011)\citenamefont
  {M\o{}lmer}, \citenamefont {Isenhower},\ and\ \citenamefont
  {Saffman}}]{Molmer2011}%
  \BibitemOpen
  \bibfield  {author} {\bibinfo {author} {\bibfnamefont {K.}~\bibnamefont
  {M\o{}lmer}}, \bibinfo {author} {\bibfnamefont {L.}~\bibnamefont
  {Isenhower}}, \ and\ \bibinfo {author} {\bibfnamefont {M.}~\bibnamefont
  {Saffman}},\ }\href@noop {} {\bibfield  {journal} {\bibinfo  {journal} {J.
  Phys. B: At. Mol. Opt. Phys.}\ }\textbf {\bibinfo {volume} {44}},\ \bibinfo
  {pages} {184016} (\bibinfo {year} {2011})}\BibitemShut {NoStop}%
\bibitem [{\citenamefont {Weimer}\ \emph {et~al.}(2010)\citenamefont {Weimer},
  \citenamefont {M\"uller}, \citenamefont {Lesanovsky}, \citenamefont
  {Zoller},\ and\ \citenamefont {B\"uchler}}]{Weimer2010}%
  \BibitemOpen
  \bibfield  {author} {\bibinfo {author} {\bibfnamefont {H.}~\bibnamefont
  {Weimer}}, \bibinfo {author} {\bibfnamefont {M.}~\bibnamefont {M\"uller}},
  \bibinfo {author} {\bibfnamefont {I.}~\bibnamefont {Lesanovsky}}, \bibinfo
  {author} {\bibfnamefont {P.}~\bibnamefont {Zoller}}, \ and\ \bibinfo {author}
  {\bibfnamefont {H.~P.}\ \bibnamefont {B\"uchler}},\ }\href@noop {} {\bibfield
   {journal} {\bibinfo  {journal} {Nat. Phys.}\ }\textbf {\bibinfo {volume}
  {6}},\ \bibinfo {pages} {382} (\bibinfo {year} {2010})}\BibitemShut {NoStop}%
\bibitem [{\citenamefont {Weimer}\ \emph {et~al.}(2011)\citenamefont {Weimer},
  \citenamefont {M\"uller}, \citenamefont {B\"uchler},\ and\ \citenamefont
  {Lesanovsky}}]{Weimer2011}%
  \BibitemOpen
  \bibfield  {author} {\bibinfo {author} {\bibfnamefont {H.}~\bibnamefont
  {Weimer}}, \bibinfo {author} {\bibfnamefont {M.}~\bibnamefont {M\"uller}},
  \bibinfo {author} {\bibfnamefont {H.~P.}\ \bibnamefont {B\"uchler}}, \ and\
  \bibinfo {author} {\bibfnamefont {I.}~\bibnamefont {Lesanovsky}},\
  }\href@noop {} {\bibfield  {journal} {\bibinfo  {journal} {Quant. Inf.
  Proc.}\ }\textbf {\bibinfo {volume} {10}},\ \bibinfo {pages} {885} (\bibinfo
  {year} {2011})}\BibitemShut {NoStop}%
\bibitem [{\citenamefont {Han}\ \emph {et~al.}(2010)\citenamefont {Han},
  \citenamefont {He}, \citenamefont {Heshami}, \citenamefont {Li},\ and\
  \citenamefont {Simon}}]{Han2010}%
  \BibitemOpen
  \bibfield  {author} {\bibinfo {author} {\bibfnamefont {Y.}~\bibnamefont
  {Han}}, \bibinfo {author} {\bibfnamefont {B.}~\bibnamefont {He}}, \bibinfo
  {author} {\bibfnamefont {K.}~\bibnamefont {Heshami}}, \bibinfo {author}
  {\bibfnamefont {C.-Z.}\ \bibnamefont {Li}}, \ and\ \bibinfo {author}
  {\bibfnamefont {C.}~\bibnamefont {Simon}},\ }\href@noop {} {\bibfield
  {journal} {\bibinfo  {journal} {Phys. Rev. A}\ }\textbf {\bibinfo {volume}
  {81}},\ \bibinfo {pages} {052311} (\bibinfo {year} {2010})}\BibitemShut
  {NoStop}%
\bibitem [{\citenamefont {Zhao}\ \emph {et~al.}(2010)\citenamefont {Zhao},
  \citenamefont {M\"uller}, \citenamefont {Hammerer},\ and\ \citenamefont
  {Zoller}}]{Zhao2010}%
  \BibitemOpen
  \bibfield  {author} {\bibinfo {author} {\bibfnamefont {B.}~\bibnamefont
  {Zhao}}, \bibinfo {author} {\bibfnamefont {M.}~\bibnamefont {M\"uller}},
  \bibinfo {author} {\bibfnamefont {K.}~\bibnamefont {Hammerer}}, \ and\
  \bibinfo {author} {\bibfnamefont {P.}~\bibnamefont {Zoller}},\ }\href@noop {}
  {\bibfield  {journal} {\bibinfo  {journal} {Phys. Rev. A}\ }\textbf {\bibinfo
  {volume} {81}},\ \bibinfo {pages} {052329} (\bibinfo {year}
  {2010})}\BibitemShut {NoStop}%
\bibitem [{\citenamefont {Urban}\ \emph {et~al.}(2009)\citenamefont {Urban},
  \citenamefont {Johnson}, \citenamefont {Henage}, \citenamefont {Isenhower},
  \citenamefont {Yavuz}, \citenamefont {Walker},\ and\ \citenamefont
  {Saffman}}]{Urban2009}%
  \BibitemOpen
  \bibfield  {author} {\bibinfo {author} {\bibfnamefont {E.}~\bibnamefont
  {Urban}}, \bibinfo {author} {\bibfnamefont {T.~A.}\ \bibnamefont {Johnson}},
  \bibinfo {author} {\bibfnamefont {T.}~\bibnamefont {Henage}}, \bibinfo
  {author} {\bibfnamefont {L.}~\bibnamefont {Isenhower}}, \bibinfo {author}
  {\bibfnamefont {D.~D.}\ \bibnamefont {Yavuz}}, \bibinfo {author}
  {\bibfnamefont {T.~G.}\ \bibnamefont {Walker}}, \ and\ \bibinfo {author}
  {\bibfnamefont {M.}~\bibnamefont {Saffman}},\ }\href@noop {} {\bibfield
  {journal} {\bibinfo  {journal} {Nature Phys.}\ }\textbf {\bibinfo {volume}
  {5}},\ \bibinfo {pages} {110} (\bibinfo {year} {2009})}\BibitemShut {NoStop}%
\bibitem [{\citenamefont {Ga\"etan}\ \emph {et~al.}(2009)\citenamefont
  {Ga\"etan}, \citenamefont {Miroshnychenko}, \citenamefont {Wilk},
  \citenamefont {Chotia}, \citenamefont {Viteau}, \citenamefont {Comparat},
  \citenamefont {Pillet}, \citenamefont {Browaeys},\ and\ \citenamefont
  {Grangier}}]{Gaetan2009}%
  \BibitemOpen
  \bibfield  {author} {\bibinfo {author} {\bibfnamefont {A.}~\bibnamefont
  {Ga\"etan}}, \bibinfo {author} {\bibfnamefont {Y.}~\bibnamefont
  {Miroshnychenko}}, \bibinfo {author} {\bibfnamefont {T.}~\bibnamefont
  {Wilk}}, \bibinfo {author} {\bibfnamefont {A.}~\bibnamefont {Chotia}},
  \bibinfo {author} {\bibfnamefont {M.}~\bibnamefont {Viteau}}, \bibinfo
  {author} {\bibfnamefont {D.}~\bibnamefont {Comparat}}, \bibinfo {author}
  {\bibfnamefont {P.}~\bibnamefont {Pillet}}, \bibinfo {author} {\bibfnamefont
  {A.}~\bibnamefont {Browaeys}}, \ and\ \bibinfo {author} {\bibfnamefont
  {P.}~\bibnamefont {Grangier}},\ }\href@noop {} {\bibfield  {journal}
  {\bibinfo  {journal} {Nature Phys.}\ }\textbf {\bibinfo {volume} {5}},\
  \bibinfo {pages} {115} (\bibinfo {year} {2009})}\BibitemShut {NoStop}%
\bibitem [{\citenamefont {Isenhower}\ \emph {et~al.}(2010)\citenamefont
  {Isenhower}, \citenamefont {Urban}, \citenamefont {Zhang}, \citenamefont
  {Gill}, \citenamefont {Henage}, \citenamefont {Johnson}, \citenamefont
  {Walker},\ and\ \citenamefont {Saffman}}]{Isenhower2010}%
  \BibitemOpen
  \bibfield  {author} {\bibinfo {author} {\bibfnamefont {L.}~\bibnamefont
  {Isenhower}}, \bibinfo {author} {\bibfnamefont {E.}~\bibnamefont {Urban}},
  \bibinfo {author} {\bibfnamefont {X.~L.}\ \bibnamefont {Zhang}}, \bibinfo
  {author} {\bibfnamefont {A.~T.}\ \bibnamefont {Gill}}, \bibinfo {author}
  {\bibfnamefont {T.}~\bibnamefont {Henage}}, \bibinfo {author} {\bibfnamefont
  {T.~A.}\ \bibnamefont {Johnson}}, \bibinfo {author} {\bibfnamefont {T.~G.}\
  \bibnamefont {Walker}}, \ and\ \bibinfo {author} {\bibfnamefont
  {M.}~\bibnamefont {Saffman}},\ }\href@noop {} {\bibfield  {journal} {\bibinfo
   {journal} {Phys. Rev. Lett.}\ }\textbf {\bibinfo {volume} {104}},\ \bibinfo
  {pages} {010503} (\bibinfo {year} {2010})}\BibitemShut {NoStop}%
\bibitem [{\citenamefont {Wilk}\ \emph {et~al.}(2010)\citenamefont {Wilk},
  \citenamefont {Ga\"etan}, \citenamefont {Evellin}, \citenamefont {Wolters},
  \citenamefont {Miroshnychenko}, \citenamefont {Grangier},\ and\ \citenamefont
  {Browaeys}}]{Wilk2010}%
  \BibitemOpen
  \bibfield  {author} {\bibinfo {author} {\bibfnamefont {T.}~\bibnamefont
  {Wilk}}, \bibinfo {author} {\bibfnamefont {A.}~\bibnamefont {Ga\"etan}},
  \bibinfo {author} {\bibfnamefont {C.}~\bibnamefont {Evellin}}, \bibinfo
  {author} {\bibfnamefont {J.}~\bibnamefont {Wolters}}, \bibinfo {author}
  {\bibfnamefont {Y.}~\bibnamefont {Miroshnychenko}}, \bibinfo {author}
  {\bibfnamefont {P.}~\bibnamefont {Grangier}}, \ and\ \bibinfo {author}
  {\bibfnamefont {A.}~\bibnamefont {Browaeys}},\ }\href@noop {} {\bibfield
  {journal} {\bibinfo  {journal} {Phys. Rev. Lett.}\ }\textbf {\bibinfo
  {volume} {104}},\ \bibinfo {pages} {010502} (\bibinfo {year}
  {2010})}\BibitemShut {NoStop}%
\bibitem [{\citenamefont {Ga\"etan}\ \emph {et~al.}(2010)\citenamefont
  {Ga\"etan}, \citenamefont {Evellin}, \citenamefont {Wolters}, \citenamefont
  {Grangier}, \citenamefont {Wilk},\ and\ \citenamefont
  {Browaeys}}]{Gaetan2010}%
  \BibitemOpen
  \bibfield  {author} {\bibinfo {author} {\bibfnamefont {A.}~\bibnamefont
  {Ga\"etan}}, \bibinfo {author} {\bibfnamefont {C.}~\bibnamefont {Evellin}},
  \bibinfo {author} {\bibfnamefont {J.}~\bibnamefont {Wolters}}, \bibinfo
  {author} {\bibfnamefont {P.}~\bibnamefont {Grangier}}, \bibinfo {author}
  {\bibfnamefont {T.}~\bibnamefont {Wilk}}, \ and\ \bibinfo {author}
  {\bibfnamefont {A.}~\bibnamefont {Browaeys}},\ }\href@noop {} {\bibfield
  {journal} {\bibinfo  {journal} {New J. Phys.}\ }\textbf {\bibinfo {volume}
  {12}},\ \bibinfo {pages} {065040} (\bibinfo {year} {2010})}\BibitemShut
  {NoStop}%
\bibitem [{\citenamefont {Zhang}\ \emph {et~al.}(2010)\citenamefont {Zhang},
  \citenamefont {Isenhower}, \citenamefont {Gill}, \citenamefont {Walker},\
  and\ \citenamefont {Saffman}}]{Zhang2010}%
  \BibitemOpen
  \bibfield  {author} {\bibinfo {author} {\bibfnamefont {X.~L.}\ \bibnamefont
  {Zhang}}, \bibinfo {author} {\bibfnamefont {L.}~\bibnamefont {Isenhower}},
  \bibinfo {author} {\bibfnamefont {A.~T.}\ \bibnamefont {Gill}}, \bibinfo
  {author} {\bibfnamefont {T.~G.}\ \bibnamefont {Walker}}, \ and\ \bibinfo
  {author} {\bibfnamefont {M.}~\bibnamefont {Saffman}},\ }\href@noop {}
  {\bibfield  {journal} {\bibinfo  {journal} {Phys. Rev. A}\ }\textbf {\bibinfo
  {volume} {82}},\ \bibinfo {pages} {030306(R)} (\bibinfo {year}
  {2010})}\BibitemShut {NoStop}%
\bibitem [{\citenamefont {Saffman}\ and\ \citenamefont
  {Walker}(2005{\natexlab{b}})}]{Saffman2005a}%
  \BibitemOpen
  \bibfield  {author} {\bibinfo {author} {\bibfnamefont {M.}~\bibnamefont
  {Saffman}}\ and\ \bibinfo {author} {\bibfnamefont {T.~G.}\ \bibnamefont
  {Walker}},\ }\href@noop {} {\bibfield  {journal} {\bibinfo  {journal} {Phys.
  Rev. A}\ }\textbf {\bibinfo {volume} {72}},\ \bibinfo {pages} {022347}
  (\bibinfo {year} {2005}{\natexlab{b}})}\BibitemShut {NoStop}%
\bibitem [{\citenamefont {Zhang}\ \emph
  {et~al.}(2012{\natexlab{a}})\citenamefont {Zhang}, \citenamefont {Gill},
  \citenamefont {Isenhower}, \citenamefont {Walker},\ and\ \citenamefont
  {Saffman}}]{XZhang2012}%
  \BibitemOpen
  \bibfield  {author} {\bibinfo {author} {\bibfnamefont {X.~L.}\ \bibnamefont
  {Zhang}}, \bibinfo {author} {\bibfnamefont {A.~T.}\ \bibnamefont {Gill}},
  \bibinfo {author} {\bibfnamefont {L.}~\bibnamefont {Isenhower}}, \bibinfo
  {author} {\bibfnamefont {T.~G.}\ \bibnamefont {Walker}}, \ and\ \bibinfo
  {author} {\bibfnamefont {M.}~\bibnamefont {Saffman}},\ }\href@noop {}
  {\bibfield  {journal} {\bibinfo  {journal} {Phys. Rev. A}\ }\textbf {\bibinfo
  {volume} {85}},\ \bibinfo {pages} {042310} (\bibinfo {year}
  {2012}{\natexlab{a}})}\BibitemShut {NoStop}%
\bibitem [{\citenamefont {Aliferis}\ \emph {et~al.}(2006)\citenamefont
  {Aliferis}, \citenamefont {Gottesman},\ and\ \citenamefont
  {Preskill}}]{Aliferis2006}%
  \BibitemOpen
  \bibfield  {author} {\bibinfo {author} {\bibfnamefont {P.}~\bibnamefont
  {Aliferis}}, \bibinfo {author} {\bibfnamefont {D.}~\bibnamefont {Gottesman}},
  \ and\ \bibinfo {author} {\bibfnamefont {J.}~\bibnamefont {Preskill}},\
  }\href@noop {} {\bibfield  {journal} {\bibinfo  {journal} {Qu. Inf. Comp.}\
  }\textbf {\bibinfo {volume} {6}},\ \bibinfo {pages} {97} (\bibinfo {year}
  {2006})}\BibitemShut {NoStop}%
\bibitem [{\citenamefont {Aliferis}\ and\ \citenamefont
  {Preskill}(2009)}]{Aliferis2009}%
  \BibitemOpen
  \bibfield  {author} {\bibinfo {author} {\bibfnamefont {P.}~\bibnamefont
  {Aliferis}}\ and\ \bibinfo {author} {\bibfnamefont {J.}~\bibnamefont
  {Preskill}},\ }\href@noop {} {\bibfield  {journal} {\bibinfo  {journal}
  {Phys. Rev. A}\ }\textbf {\bibinfo {volume} {79}},\ \bibinfo {pages} {012332}
  (\bibinfo {year} {2009})}\BibitemShut {NoStop}%
\bibitem [{\citenamefont {Fowler}\ \emph {et~al.}(2009)\citenamefont {Fowler},
  \citenamefont {Stephens},\ and\ \citenamefont {Groszkowski}}]{Fowler2009}%
  \BibitemOpen
  \bibfield  {author} {\bibinfo {author} {\bibfnamefont {A.~G.}\ \bibnamefont
  {Fowler}}, \bibinfo {author} {\bibfnamefont {A.~M.}\ \bibnamefont
  {Stephens}}, \ and\ \bibinfo {author} {\bibfnamefont {P.}~\bibnamefont
  {Groszkowski}},\ }\href@noop {} {\bibfield  {journal} {\bibinfo  {journal}
  {Phys. Rev. A}\ }\textbf {\bibinfo {volume} {80}},\ \bibinfo {pages} {052312}
  (\bibinfo {year} {2009})}\BibitemShut {NoStop}%
\bibitem [{\citenamefont {Beterov}\ \emph {et~al.}(2011)\citenamefont
  {Beterov}, \citenamefont {Tretyakov}, \citenamefont {Entin}, \citenamefont
  {Yakshina}, \citenamefont {Ryabtsev}, \citenamefont {MacCormick},\ and\
  \citenamefont {Bergamini}}]{Beterov2011}%
  \BibitemOpen
  \bibfield  {author} {\bibinfo {author} {\bibfnamefont {I.~I.}\ \bibnamefont
  {Beterov}}, \bibinfo {author} {\bibfnamefont {D.~B.}\ \bibnamefont
  {Tretyakov}}, \bibinfo {author} {\bibfnamefont {V.~M.}\ \bibnamefont
  {Entin}}, \bibinfo {author} {\bibfnamefont {E.~A.}\ \bibnamefont {Yakshina}},
  \bibinfo {author} {\bibfnamefont {I.~I.}\ \bibnamefont {Ryabtsev}}, \bibinfo
  {author} {\bibfnamefont {C.}~\bibnamefont {MacCormick}}, \ and\ \bibinfo
  {author} {\bibfnamefont {S.}~\bibnamefont {Bergamini}},\ }\href@noop {}
  {\bibfield  {journal} {\bibinfo  {journal} {Phys. Rev. A}\ }\textbf {\bibinfo
  {volume} {84}},\ \bibinfo {pages} {023413} (\bibinfo {year}
  {2011})}\BibitemShut {NoStop}%
\bibitem [{\citenamefont {Zhang}\ \emph {et~al.}(2011)\citenamefont {Zhang},
  \citenamefont {Robicheaux},\ and\ \citenamefont {Saffman}}]{SZhang2011}%
  \BibitemOpen
  \bibfield  {author} {\bibinfo {author} {\bibfnamefont {S.}~\bibnamefont
  {Zhang}}, \bibinfo {author} {\bibfnamefont {F.}~\bibnamefont {Robicheaux}}, \
  and\ \bibinfo {author} {\bibfnamefont {M.}~\bibnamefont {Saffman}},\
  }\href@noop {} {\bibfield  {journal} {\bibinfo  {journal} {Phys. Rev. A}\
  }\textbf {\bibinfo {volume} {84}},\ \bibinfo {pages} {043408} (\bibinfo
  {year} {2011})}\BibitemShut {NoStop}%
\bibitem [{\citenamefont {Morrison}\ and\ \citenamefont
  {Derevianko}(2012)}]{Morrison2012}%
  \BibitemOpen
  \bibfield  {author} {\bibinfo {author} {\bibfnamefont {M.~J.}\ \bibnamefont
  {Morrison}}\ and\ \bibinfo {author} {\bibfnamefont {A.}~\bibnamefont
  {Derevianko}},\ }\href@noop {} {\bibfield  {journal} {\bibinfo  {journal}
  {Phys. Rev. A}\ }\textbf {\bibinfo {volume} {85}},\ \bibinfo {pages} {033414}
  (\bibinfo {year} {2012})}\BibitemShut {NoStop}%
\bibitem [{\citenamefont {Chuang}\ and\ \citenamefont
  {Nielsen}(1997)}]{Chuang1997}%
  \BibitemOpen
  \bibfield  {author} {\bibinfo {author} {\bibfnamefont {I.~L.}\ \bibnamefont
  {Chuang}}\ and\ \bibinfo {author} {\bibfnamefont {M.~A.}\ \bibnamefont
  {Nielsen}},\ }\href@noop {} {\bibfield  {journal} {\bibinfo  {journal} {J.
  Mod. Opt.}\ }\textbf {\bibinfo {volume} {44}},\ \bibinfo {pages} {2455}
  (\bibinfo {year} {1997})}\BibitemShut {NoStop}%
\bibitem [{\citenamefont {Poyatos}\ \emph {et~al.}(1997)\citenamefont
  {Poyatos}, \citenamefont {Cirac},\ and\ \citenamefont
  {Zoller}}]{Poyatos1997}%
  \BibitemOpen
  \bibfield  {author} {\bibinfo {author} {\bibfnamefont {J.~F.}\ \bibnamefont
  {Poyatos}}, \bibinfo {author} {\bibfnamefont {J.~I.}\ \bibnamefont {Cirac}},
  \ and\ \bibinfo {author} {\bibfnamefont {P.}~\bibnamefont {Zoller}},\
  }\href@noop {} {\bibfield  {journal} {\bibinfo  {journal} {Phys. Rev. Lett.}\
  }\textbf {\bibinfo {volume} {78}},\ \bibinfo {pages} {390} (\bibinfo {year}
  {1997})}\BibitemShut {NoStop}%
\bibitem [{\citenamefont {Nielsen}\ and\ \citenamefont
  {Chuang}(2000)}]{Nielsen2000}%
  \BibitemOpen
  \bibfield  {author} {\bibinfo {author} {\bibfnamefont {M.~A.}\ \bibnamefont
  {Nielsen}}\ and\ \bibinfo {author} {\bibfnamefont {I.~L.}\ \bibnamefont
  {Chuang}},\ }\href@noop {} {\emph {\bibinfo {title} {Quantum computation and
  quantum information}}}\ (\bibinfo  {publisher} {Cambridge University Press,
  Cambridge},\ \bibinfo {year} {2000})\BibitemShut {NoStop}%
\bibitem [{\citenamefont {O'Brien}\ \emph {et~al.}(2004)\citenamefont
  {O'Brien}, \citenamefont {Pryde}, \citenamefont {Gilchrist}, \citenamefont
  {James}, \citenamefont {Langford}, \citenamefont {Ralph},\ and\ \citenamefont
  {White}}]{OBrien2004}%
  \BibitemOpen
  \bibfield  {author} {\bibinfo {author} {\bibfnamefont {J.~L.}\ \bibnamefont
  {O'Brien}}, \bibinfo {author} {\bibfnamefont {G.~J.}\ \bibnamefont {Pryde}},
  \bibinfo {author} {\bibfnamefont {A.}~\bibnamefont {Gilchrist}}, \bibinfo
  {author} {\bibfnamefont {D.~F.~V.}\ \bibnamefont {James}}, \bibinfo {author}
  {\bibfnamefont {N.~K.}\ \bibnamefont {Langford}}, \bibinfo {author}
  {\bibfnamefont {T.~C.}\ \bibnamefont {Ralph}}, \ and\ \bibinfo {author}
  {\bibfnamefont {A.~G.}\ \bibnamefont {White}},\ }\href@noop {} {\bibfield
  {journal} {\bibinfo  {journal} {Phys. Rev. Lett.}\ }\textbf {\bibinfo
  {volume} {93}},\ \bibinfo {pages} {080502} (\bibinfo {year}
  {2004})}\BibitemShut {NoStop}%
\bibitem [{\citenamefont {White}\ \emph {et~al.}(2007)\citenamefont {White},
  \citenamefont {Gilchrist}, \citenamefont {Pryde}, \citenamefont {O'Brien},
  \citenamefont {Bremner},\ and\ \citenamefont {Langford}}]{White2007}%
  \BibitemOpen
  \bibfield  {author} {\bibinfo {author} {\bibfnamefont {A.~G.}\ \bibnamefont
  {White}}, \bibinfo {author} {\bibfnamefont {A.}~\bibnamefont {Gilchrist}},
  \bibinfo {author} {\bibfnamefont {G.~J.}\ \bibnamefont {Pryde}}, \bibinfo
  {author} {\bibfnamefont {J.~L.}\ \bibnamefont {O'Brien}}, \bibinfo {author}
  {\bibfnamefont {M.~J.}\ \bibnamefont {Bremner}}, \ and\ \bibinfo {author}
  {\bibfnamefont {N.~K.}\ \bibnamefont {Langford}},\ }\href@noop {} {\bibfield
  {journal} {\bibinfo  {journal} {J. Opt. Soc. Am. B}\ }\textbf {\bibinfo
  {volume} {24}},\ \bibinfo {pages} {172} (\bibinfo {year} {2007})}\BibitemShut
  {NoStop}%
\bibitem [{\citenamefont {Riebe}\ \emph {et~al.}(2006)\citenamefont {Riebe},
  \citenamefont {Kim}, \citenamefont {Schindler}, \citenamefont {Monz},
  \citenamefont {Schmidt}, \citenamefont {K\"orber}, \citenamefont {H\"ansel},
  \citenamefont {H\"affner}, \citenamefont {Roos},\ and\ \citenamefont
  {Blatt}}]{Riebe2006}%
  \BibitemOpen
  \bibfield  {author} {\bibinfo {author} {\bibfnamefont {M.}~\bibnamefont
  {Riebe}}, \bibinfo {author} {\bibfnamefont {K.}~\bibnamefont {Kim}}, \bibinfo
  {author} {\bibfnamefont {P.}~\bibnamefont {Schindler}}, \bibinfo {author}
  {\bibfnamefont {T.}~\bibnamefont {Monz}}, \bibinfo {author} {\bibfnamefont
  {P.~O.}\ \bibnamefont {Schmidt}}, \bibinfo {author} {\bibfnamefont {T.~K.}\
  \bibnamefont {K\"orber}}, \bibinfo {author} {\bibfnamefont {W.}~\bibnamefont
  {H\"ansel}}, \bibinfo {author} {\bibfnamefont {H.}~\bibnamefont {H\"affner}},
  \bibinfo {author} {\bibfnamefont {C.~F.}\ \bibnamefont {Roos}}, \ and\
  \bibinfo {author} {\bibfnamefont {R.}~\bibnamefont {Blatt}},\ }\href@noop {}
  {\bibfield  {journal} {\bibinfo  {journal} {Phys. Rev. Lett.}\ }\textbf
  {\bibinfo {volume} {97}},\ \bibinfo {pages} {220407} (\bibinfo {year}
  {2006})}\BibitemShut {NoStop}%
\bibitem [{\citenamefont {Wang}\ \emph {et~al.}(2010)\citenamefont {Wang},
  \citenamefont {Labaziewicz}, \citenamefont {Ge}, \citenamefont {Shewmon},\
  and\ \citenamefont {Chuang}}]{SXWang2010}%
  \BibitemOpen
  \bibfield  {author} {\bibinfo {author} {\bibfnamefont {S.~X.}\ \bibnamefont
  {Wang}}, \bibinfo {author} {\bibfnamefont {J.}~\bibnamefont {Labaziewicz}},
  \bibinfo {author} {\bibfnamefont {Y.}~\bibnamefont {Ge}}, \bibinfo {author}
  {\bibfnamefont {R.}~\bibnamefont {Shewmon}}, \ and\ \bibinfo {author}
  {\bibfnamefont {I.~L.}\ \bibnamefont {Chuang}},\ }\href@noop {} {\bibfield
  {journal} {\bibinfo  {journal} {Phys. Rev. A}\ }\textbf {\bibinfo {volume}
  {81}},\ \bibinfo {pages} {062332} (\bibinfo {year} {2010})}\BibitemShut
  {NoStop}%
\bibitem [{\citenamefont {Yamamoto}\ \emph {et~al.}(2010)\citenamefont
  {Yamamoto}, \citenamefont {Neeley}, \citenamefont {Lucero}, \citenamefont
  {Bialczak}, \citenamefont {Kelly}, \citenamefont {Lenander}, \citenamefont
  {Mariantoni}, \citenamefont {O'Connell}, \citenamefont {Sank}, \citenamefont
  {Wang}, \citenamefont {Weides}, \citenamefont {Wenner}, \citenamefont {Yin},
  \citenamefont {Cleland},\ and\ \citenamefont {Martinis}}]{Yamamoto2010}%
  \BibitemOpen
  \bibfield  {author} {\bibinfo {author} {\bibfnamefont {T.}~\bibnamefont
  {Yamamoto}}, \bibinfo {author} {\bibfnamefont {M.}~\bibnamefont {Neeley}},
  \bibinfo {author} {\bibfnamefont {E.}~\bibnamefont {Lucero}}, \bibinfo
  {author} {\bibfnamefont {R.~C.}\ \bibnamefont {Bialczak}}, \bibinfo {author}
  {\bibfnamefont {J.}~\bibnamefont {Kelly}}, \bibinfo {author} {\bibfnamefont
  {M.}~\bibnamefont {Lenander}}, \bibinfo {author} {\bibfnamefont
  {M.}~\bibnamefont {Mariantoni}}, \bibinfo {author} {\bibfnamefont {A.~D.}\
  \bibnamefont {O'Connell}}, \bibinfo {author} {\bibfnamefont {D.}~\bibnamefont
  {Sank}}, \bibinfo {author} {\bibfnamefont {H.}~\bibnamefont {Wang}}, \bibinfo
  {author} {\bibfnamefont {M.}~\bibnamefont {Weides}}, \bibinfo {author}
  {\bibfnamefont {J.}~\bibnamefont {Wenner}}, \bibinfo {author} {\bibfnamefont
  {Y.}~\bibnamefont {Yin}}, \bibinfo {author} {\bibfnamefont {A.~N.}\
  \bibnamefont {Cleland}}, \ and\ \bibinfo {author} {\bibfnamefont {J.~M.}\
  \bibnamefont {Martinis}},\ }\href@noop {} {\bibfield  {journal} {\bibinfo
  {journal} {Phys. Rev. B}\ }\textbf {\bibinfo {volume} {82}},\ \bibinfo
  {pages} {184515} (\bibinfo {year} {2010})}\BibitemShut {NoStop}%
\bibitem [{\citenamefont {Bialczak}\ \emph {et~al.}(2010)\citenamefont
  {Bialczak}, \citenamefont {Ansmann}, \citenamefont {Hofheinz}, \citenamefont
  {Lucero}, \citenamefont {Neeley}, \citenamefont {O'Connell}, \citenamefont
  {Sank}, \citenamefont {Wang}, \citenamefont {Wenner}, \citenamefont
  {Steffen}, \citenamefont {Cleland},\ and\ \citenamefont
  {Martinis}}]{Bialczak2010}%
  \BibitemOpen
  \bibfield  {author} {\bibinfo {author} {\bibfnamefont {R.~C.}\ \bibnamefont
  {Bialczak}}, \bibinfo {author} {\bibfnamefont {M.}~\bibnamefont {Ansmann}},
  \bibinfo {author} {\bibfnamefont {M.}~\bibnamefont {Hofheinz}}, \bibinfo
  {author} {\bibfnamefont {E.}~\bibnamefont {Lucero}}, \bibinfo {author}
  {\bibfnamefont {M.}~\bibnamefont {Neeley}}, \bibinfo {author} {\bibfnamefont
  {A.~D.}\ \bibnamefont {O'Connell}}, \bibinfo {author} {\bibfnamefont
  {D.}~\bibnamefont {Sank}}, \bibinfo {author} {\bibfnamefont {H.}~\bibnamefont
  {Wang}}, \bibinfo {author} {\bibfnamefont {J.}~\bibnamefont {Wenner}},
  \bibinfo {author} {\bibfnamefont {M.}~\bibnamefont {Steffen}}, \bibinfo
  {author} {\bibfnamefont {A.~N.}\ \bibnamefont {Cleland}}, \ and\ \bibinfo
  {author} {\bibfnamefont {J.~M.}\ \bibnamefont {Martinis}},\ }\href@noop {}
  {\bibfield  {journal} {\bibinfo  {journal} {Nat. Phys.}\ }\textbf {\bibinfo
  {volume} {6}},\ \bibinfo {pages} {409} (\bibinfo {year} {2010})}\BibitemShut
  {NoStop}%
\bibitem [{\citenamefont {Zhang}\ \emph
  {et~al.}(2012{\natexlab{b}})\citenamefont {Zhang}, \citenamefont {Hung},
  \citenamefont {Tung},\ and\ \citenamefont {Chin}}]{Zhang2012}%
  \BibitemOpen
  \bibfield  {author} {\bibinfo {author} {\bibfnamefont {X.}~\bibnamefont
  {Zhang}}, \bibinfo {author} {\bibfnamefont {C.-L.}\ \bibnamefont {Hung}},
  \bibinfo {author} {\bibfnamefont {S.-K.}\ \bibnamefont {Tung}}, \ and\
  \bibinfo {author} {\bibfnamefont {C.}~\bibnamefont {Chin}},\ }\href@noop {}
  {\bibfield  {journal} {\bibinfo  {journal} {Science}\ }\textbf {\bibinfo
  {volume} {335}},\ \bibinfo {pages} {1070} (\bibinfo {year}
  {2012}{\natexlab{b}})}\BibitemShut {NoStop}%
\bibitem [{\citenamefont {Nielsen}(2002)}]{Nielsen2002}%
  \BibitemOpen
  \bibfield  {author} {\bibinfo {author} {\bibfnamefont {M.~A.}\ \bibnamefont
  {Nielsen}},\ }\href@noop {} {\bibfield  {journal} {\bibinfo  {journal} {Phys.
  Lett. A}\ }\textbf {\bibinfo {volume} {303}},\ \bibinfo {pages} {249}
  (\bibinfo {year} {2002})}\BibitemShut {NoStop}%
\bibitem [{\citenamefont {Pedersen}\ \emph {et~al.}(2007)\citenamefont
  {Pedersen}, \citenamefont {M\o{}ller},\ and\ \citenamefont
  {M\o{}lmer}}]{Pedersen2007}%
  \BibitemOpen
  \bibfield  {author} {\bibinfo {author} {\bibfnamefont {L.~H.}\ \bibnamefont
  {Pedersen}}, \bibinfo {author} {\bibfnamefont {N.~M.}\ \bibnamefont
  {M\o{}ller}}, \ and\ \bibinfo {author} {\bibfnamefont {K.}~\bibnamefont
  {M\o{}lmer}},\ }\href@noop {} {\bibfield  {journal} {\bibinfo  {journal}
  {Phys. Lett. A}\ }\textbf {\bibinfo {volume} {367}},\ \bibinfo {pages} {47}
  (\bibinfo {year} {2007})}\BibitemShut {NoStop}%
\bibitem [{\citenamefont {Hulet}\ and\ \citenamefont
  {Kleppner}(1983)}]{Hulet1983}%
  \BibitemOpen
  \bibfield  {author} {\bibinfo {author} {\bibfnamefont {R.~G.}\ \bibnamefont
  {Hulet}}\ and\ \bibinfo {author} {\bibfnamefont {D.}~\bibnamefont
  {Kleppner}},\ }\href@noop {} {\bibfield  {journal} {\bibinfo  {journal}
  {Phys. Rev. Lett.}\ }\textbf {\bibinfo {volume} {51}},\ \bibinfo {pages}
  {1430} (\bibinfo {year} {1983})}\BibitemShut {NoStop}%
\bibitem [{\citenamefont {Hare}\ \emph {et~al.}(1988)\citenamefont {Hare},
  \citenamefont {Gross},\ and\ \citenamefont {Goy}}]{Hare1988}%
  \BibitemOpen
  \bibfield  {author} {\bibinfo {author} {\bibfnamefont {J.}~\bibnamefont
  {Hare}}, \bibinfo {author} {\bibfnamefont {M.}~\bibnamefont {Gross}}, \ and\
  \bibinfo {author} {\bibfnamefont {P.}~\bibnamefont {Goy}},\ }\href@noop {}
  {\bibfield  {journal} {\bibinfo  {journal} {Phys. Rev. Lett.}\ }\textbf
  {\bibinfo {volume} {61}},\ \bibinfo {pages} {1938} (\bibinfo {year}
  {1988})}\BibitemShut {NoStop}%
\bibitem [{\citenamefont {Nussenzweig}\ \emph {et~al.}(1991)\citenamefont
  {Nussenzweig}, \citenamefont {Hare}, \citenamefont {Steinberg}, \citenamefont
  {Moi}, \citenamefont {Gross},\ and\ \citenamefont
  {Haroche}}]{Nussenzweig1991}%
  \BibitemOpen
  \bibfield  {author} {\bibinfo {author} {\bibfnamefont {A.}~\bibnamefont
  {Nussenzweig}}, \bibinfo {author} {\bibfnamefont {J.}~\bibnamefont {Hare}},
  \bibinfo {author} {\bibfnamefont {A.~M.}\ \bibnamefont {Steinberg}}, \bibinfo
  {author} {\bibfnamefont {L.}~\bibnamefont {Moi}}, \bibinfo {author}
  {\bibfnamefont {M.}~\bibnamefont {Gross}}, \ and\ \bibinfo {author}
  {\bibfnamefont {S.}~\bibnamefont {Haroche}},\ }\href@noop {} {\bibfield
  {journal} {\bibinfo  {journal} {Europhys. Lett.}\ }\textbf {\bibinfo {volume}
  {14}},\ \bibinfo {pages} {755} (\bibinfo {year} {1991})}\BibitemShut
  {NoStop}%
\bibitem [{\citenamefont {Brecha}\ \emph {et~al.}(1993)\citenamefont {Brecha},
  \citenamefont {Raithel}, \citenamefont {Wagner},\ and\ \citenamefont
  {Walther}}]{Brecha1993}%
  \BibitemOpen
  \bibfield  {author} {\bibinfo {author} {\bibfnamefont {R.~J.}\ \bibnamefont
  {Brecha}}, \bibinfo {author} {\bibfnamefont {G.}~\bibnamefont {Raithel}},
  \bibinfo {author} {\bibfnamefont {C.}~\bibnamefont {Wagner}}, \ and\ \bibinfo
  {author} {\bibfnamefont {H.}~\bibnamefont {Walther}},\ }\href@noop {}
  {\bibfield  {journal} {\bibinfo  {journal} {Opt. Commun.}\ }\textbf {\bibinfo
  {volume} {102}},\ \bibinfo {pages} {257} (\bibinfo {year}
  {1993})}\BibitemShut {NoStop}%
\bibitem [{\citenamefont {Nussenzveig}\ \emph {et~al.}(1993)\citenamefont
  {Nussenzveig}, \citenamefont {Bernardot}, \citenamefont {Brune},
  \citenamefont {Hare}, \citenamefont {Raimond}, \citenamefont {Haroche},\ and\
  \citenamefont {Gawlik}}]{Nussenzweig1993}%
  \BibitemOpen
  \bibfield  {author} {\bibinfo {author} {\bibfnamefont {P.}~\bibnamefont
  {Nussenzveig}}, \bibinfo {author} {\bibfnamefont {F.}~\bibnamefont
  {Bernardot}}, \bibinfo {author} {\bibfnamefont {M.}~\bibnamefont {Brune}},
  \bibinfo {author} {\bibfnamefont {J.}~\bibnamefont {Hare}}, \bibinfo {author}
  {\bibfnamefont {J.~M.}\ \bibnamefont {Raimond}}, \bibinfo {author}
  {\bibfnamefont {S.}~\bibnamefont {Haroche}}, \ and\ \bibinfo {author}
  {\bibfnamefont {W.}~\bibnamefont {Gawlik}},\ }\href@noop {} {\bibfield
  {journal} {\bibinfo  {journal} {Phys. Rev. A}\ }\textbf {\bibinfo {volume}
  {48}},\ \bibinfo {pages} {3991} (\bibinfo {year} {1993})}\BibitemShut
  {NoStop}%
\bibitem [{\citenamefont {Cheng}\ \emph {et~al.}(1994)\citenamefont {Cheng},
  \citenamefont {Lee},\ and\ \citenamefont {Gallagher}}]{Cheng1994}%
  \BibitemOpen
  \bibfield  {author} {\bibinfo {author} {\bibfnamefont {C.~H.}\ \bibnamefont
  {Cheng}}, \bibinfo {author} {\bibfnamefont {C.~Y.}\ \bibnamefont {Lee}}, \
  and\ \bibinfo {author} {\bibfnamefont {T.~F.}\ \bibnamefont {Gallagher}},\
  }\href@noop {} {\bibfield  {journal} {\bibinfo  {journal} {Phys. Rev. Lett.}\
  }\textbf {\bibinfo {volume} {73}},\ \bibinfo {pages} {3078} (\bibinfo {year}
  {1994})}\BibitemShut {NoStop}%
\bibitem [{\citenamefont {Anderson}\ \emph {et~al.}(2013)\citenamefont
  {Anderson}, \citenamefont {Schwarzkopf}, \citenamefont {Sapiro},\ and\
  \citenamefont {Raithel}}]{Anderson2013}%
  \BibitemOpen
  \bibfield  {author} {\bibinfo {author} {\bibfnamefont {D.~A.}\ \bibnamefont
  {Anderson}}, \bibinfo {author} {\bibfnamefont {A.}~\bibnamefont
  {Schwarzkopf}}, \bibinfo {author} {\bibfnamefont {R.~E.}\ \bibnamefont
  {Sapiro}}, \ and\ \bibinfo {author} {\bibfnamefont {G.}~\bibnamefont
  {Raithel}},\ }\href@noop {} {\bibfield  {journal} {\bibinfo  {journal} {Phys.
  Rev. A}\ }\textbf {\bibinfo {volume} {88}},\ \bibinfo {pages} {031401}
  (\bibinfo {year} {2013})}\BibitemShut {NoStop}%
\bibitem [{\citenamefont {Molander}\ \emph {et~al.}(1986)\citenamefont
  {Molander}, \citenamefont {C.~R.~Stroud},\ and\ \citenamefont
  {Yeazell}}]{Molander1986}%
  \BibitemOpen
  \bibfield  {author} {\bibinfo {author} {\bibfnamefont {W.~A.}\ \bibnamefont
  {Molander}}, \bibinfo {author} {\bibfnamefont {J.}~\bibnamefont
  {C.~R.~Stroud}}, \ and\ \bibinfo {author} {\bibfnamefont {J.~A.}\
  \bibnamefont {Yeazell}},\ }\href@noop {} {\bibfield  {journal} {\bibinfo
  {journal} {J. Phys. B: At. Mol. Phys.}\ }\textbf {\bibinfo {volume} {19}},\
  \bibinfo {pages} {L461} (\bibinfo {year} {1986})}\BibitemShut {NoStop}%
\bibitem [{\citenamefont {Lutwak}\ \emph {et~al.}(1997)\citenamefont {Lutwak},
  \citenamefont {Holley}, \citenamefont {Chang}, \citenamefont {Paine},
  \citenamefont {Kleppner},\ and\ \citenamefont {Ducas}}]{Lutwak1997}%
  \BibitemOpen
  \bibfield  {author} {\bibinfo {author} {\bibfnamefont {R.}~\bibnamefont
  {Lutwak}}, \bibinfo {author} {\bibfnamefont {J.}~\bibnamefont {Holley}},
  \bibinfo {author} {\bibfnamefont {P.~P.}\ \bibnamefont {Chang}}, \bibinfo
  {author} {\bibfnamefont {S.}~\bibnamefont {Paine}}, \bibinfo {author}
  {\bibfnamefont {D.}~\bibnamefont {Kleppner}}, \ and\ \bibinfo {author}
  {\bibfnamefont {T.}~\bibnamefont {Ducas}},\ }\href@noop {} {\bibfield
  {journal} {\bibinfo  {journal} {Phys. Rev. A}\ }\textbf {\bibinfo {volume}
  {56}},\ \bibinfo {pages} {1443} (\bibinfo {year} {1997})}\BibitemShut
  {NoStop}%
\bibitem [{\citenamefont {Delande}\ and\ \citenamefont
  {Gay}(1988)}]{Delande1988}%
  \BibitemOpen
  \bibfield  {author} {\bibinfo {author} {\bibfnamefont {D.}~\bibnamefont
  {Delande}}\ and\ \bibinfo {author} {\bibfnamefont {J.~C.}\ \bibnamefont
  {Gay}},\ }\href@noop {} {\bibfield  {journal} {\bibinfo  {journal} {Europhys.
  Lett.}\ }\textbf {\bibinfo {volume} {5}},\ \bibinfo {pages} {303} (\bibinfo
  {year} {1988})}\BibitemShut {NoStop}%
\bibitem [{\citenamefont {Vitanov}(1998)}]{Vitanov1998}%
  \BibitemOpen
  \bibfield  {author} {\bibinfo {author} {\bibfnamefont {N.~V.}\ \bibnamefont
  {Vitanov}},\ }\href@noop {} {\bibfield  {journal} {\bibinfo  {journal} {Phys.
  Rev. A}\ }\textbf {\bibinfo {volume} {58}},\ \bibinfo {pages} {2295}
  (\bibinfo {year} {1998})}\BibitemShut {NoStop}%
\end{thebibliography}



\end{document}